\documentclass[final,5p,times,twocolumn]{elsarticle}

\usepackage[utf8]{inputenc}
\usepackage[T1]{fontenc}

\usepackage{amsmath}

\usepackage{xfrac}

\newcommand{\minus}{\scalebox{.4}[.8]{\( - \)}}
\newcommand{\pone}{1}
\newcommand{\mone}{\minus1}
\newcommand{\zero}{0}
\newcommand{\phlf}{\sfrac{1}{2}}
\newcommand{\mhlf}{\minus\sfrac{1}{2}}

\usepackage{tikz}
\usetikzlibrary{calc}

\usepackage{enumitem}
\setitemize{noitemsep,itemsep=1pt,topsep=1pt,parsep=0pt,partopsep=0pt}

\usepackage{fancyvrb}
\fvset{
  frame=single,
  fontsize=\scriptsize,
  fontfamily=courier,
  framerule=.2mm,
  framesep=.7mm,
  numbers=left,
  xleftmargin=4mm,
  numbersep=.25mm,
  firstnumber=last
}
\input{pygments}

\newcounter{lstnocpp}
\newcounter{lstnopyt}
\newcounter{lstnofor}

\newcounter{linenocpp}
\newcounter{linenopyt}
\newcounter{linenofor}

\newcommand*\FancyVerbStartString{}
\newcommand*\FancyVerbStopString{}

\newcommand{\prog}[1]{{\rm\bf#1}}
\newcommand{\url}[1]{{#1}}

\journal{Computer Physics Communications}

\begin{document}
  \begin{frontmatter}

    \title{
      Object-oriented implementations of the MPDATA advection equation solver in~C++,~Python~and~Fortran
    }

    \author[1]{Sylwester Arabas}
    \author[1]{Dorota Jarecka}
    \author[1]{Anna Jaruga}
    \author[2]{Maciej Fijałkowski}

    \address[1]{Institute of Geophysics, Faculty of Physics, University of Warsaw}
    \address[2]{PyPy Team}

    \begin{abstract}
        Three object-oriented implementations of a prototype solver of the advection equation are introduced.
        The presented programs are based on Blitz++ (C++), NumPy (Python), and Fortran's built-in array containers.
        The solvers include an implementation of the Multidimensional Positive-Definite 
          Advective Transport Algorithm (MPDATA).
        The introduced codes exemplify how the application of 
          object-oriented programming (OOP) techniques allows to reproduce the mathematical notation 
          used in the literature within the program code.
        A discussion on the tradeoffs of the programming language choice is presented.
        The main angles of comparison are code brevity and syntax clarity
          (and hence maintainability and auditability) as well as performance.
        In the case of Python, a significant performance gain is observed when switching from the standard 
          interpreter (CPython) to the PyPy implementation of Python.
        Entire source code of all three implementations is embedded in the text and is licensed
          under the terms of the GNU GPL license. 
    \end{abstract}


  \end{frontmatter}

  \tableofcontents




  \section{Introduction}

  Object oriented programming (OOP) {\em ''has become recognised as the almost unique successful 
    paradigm for creating complex software''} \citep[][Sec.~1.3]{Press_et_al_2007}.
  It is intriguing that, while the quoted statement comes from the very book subtitled 
   {\em The Art of Scientific Computing}, hardly any (if not none) of the currently operational 
    weather and climate prediction systems - flagship examples of complex scientific software - 
    make extensive use of OOP techniques.
  Fortran has been the language of choice in oceanic \citep{Griffies_et_al_2000}, 
    weather-prediction \citep{Sundberg_2009} and Earth system \citep{Legutke_2012} modelling, 
    and none of its 20-century editions were object-oriented languages \citep[see e.g.][for discussion]{Norton_et_al_2007}.

  Application of OOP techniques in development of numerical modelling software may help to:
  \begin{enumerate}[label=(\roman*), leftmargin=*, widest=ii]
    \item{maintain modularity and separation of program logic layers (e.g. separation of
      numerical algorithms, parallelisation mechanisms, data input/output, error handling and
      the description of physical processes); and}
    \item{{\bf shorten and simplify the source code and improve its readability by reproducing within 
      the program logic the mathematical notation used in the literature}.}
  \end{enumerate}
  The first application is attainable, yet arguably cumbersome, with procedural programming.
  The latter, virtually impossible to obtain with procedural programming, is the focus of this paper.
  It also enables the compiler or library authors to relieve the user (i.e. scientific programmer)
    from hand-coding optimisations, a practice long recognised as having {\em a strong negative impact when debugging
    and maintenance are considered} \citep{Knuth_1974}.

  MPDATA \citep{Smolarkiewicz_1984} stands for Multidimensional Positive Definite Advective Transport Algorithm and is
    an example of a numerical procedure used in weather, climate and ocean simulation systems
    \citep[e.g.][respectively]{Ziemianski_et_al_2011,Abiodun_et_al_2011,Ezer_et_al_2002}.
  MPDATA is a solver for systems of advection equations of the following form:
  \begin{equation}\label{eq:adv}
    \partial_t \psi = - \nabla \cdot (\vec{v} \psi)

  \end{equation}
  that describe evolution of a scalar field $\psi$ transported by the fluid flow with velocity $\vec{v}$.
  Quoting Numerical Recipes once more, development of methods to numerically solve such problems 
    {\em ''is an art as much as a science''} \citep[][Sec.~20.1]{Press_et_al_2007},
    and MPDATA is an example of the state-of-the art in this field.
  MPDATA is designed to accurately solve equation (\ref{eq:adv}) in an arbitrary
    number of dimensions assuring positive-definiteness of scalar field $\psi$ 
    and incurring small numerical diffusion.
  All relevant MPDATA formul\ae~are given in the text but are presented without
    derivation or detailed discussion.
  For a recent review of MPDATA-based techniques see \citet[][and references therein]{Smolarkiewicz_2006}.

  In this paper we introduce and discuss object-oriented implementations of an MPDATA-based 
    two-dimensional (2D) advection equation solver written in C++11 (\citeauthor{ISO_14882} 14882:2011), 
    Python \citep{Rossum_and_Drake_2011} and Fortran 2008 (\citeauthor{ISO_1539-1} 1539-1:2010). 
  In the following section we introduce the three implementations
    briefly describing the algorithm itself and
    discussing where and how the OOP techniques may be applied in its implementation.
  The syntax and nomenclature of OOP techniques are used without introduction,
    for an overview of OOP in context of C++, Python and Fortran, consult for example
    \citep[][Part~II]{Stroustrup_2000}, \citep[][Chapter~5]{Pilgrim_2004} and
    \citep[][Chapter~11]{Markus_2012}, respectively.
  The third section of this paper covers performance evaluation of the three implementations.
  The fourth section covers discussion of the tradeoffs of the programming language choice.
  The fifth section closes the article with a brief summary.

  Throughout the paper we present the three implementations by discussing 
    source code listings which cover the entire program code.
  Subsections \ref{sec:array}-\ref{sec:solver} describe all three implementations,
    while subsequent sections \ref{sec:cyclic}-\ref{sec:example} cover discussion of C++ code 
    only.
  The relevant parts of Python and Fortran codes do not differ significantly, and for readability reasons 
    are presented in \ref{app:P} and \ref{app:F}, respectively.

  The entire code is licensed under the terms of the GNU General Public License license version 3 \citep{GPLv3}.

  All listings include line numbers printed to the left of the source code, with separate numbering for
    C++ (listings prefixed with C, black frame),
  \addtocounter{lstnocpp}{1}%
  \renewcommand*\FancyVerbStartString{\PY{c+c1}{//listing00}}%
  \renewcommand*\FancyVerbStopString{\PY{c+c1}{//listing01}}%
  \setcounter{FancyVerbLine}{\thelinenocpp}%
  \fvset{label={listing~C.\thelstnocpp~(C++)},rulecolor=\color{black},stepnumber=1}%

  \setcounter{linenocpp}{\value{FancyVerbLine}}%

    Python (listings prefixed with P, blue frame) and
  \addtocounter{lstnopyt}{1}%
  \renewcommand*\FancyVerbStartString{\PY{c}{\PYZsh{}listing00}}
  \renewcommand*\FancyVerbStopString{\PY{c}{\PYZsh{}listing01}}
  \setcounter{FancyVerbLine}{\thelinenopyt}%
  \fvset{label={listing~P.\thelstnopyt~(Python)},rulecolor=\color{blue},stepnumber=1}%
  %
  \setcounter{linenopyt}{\value{FancyVerbLine}}%

    Fortran (listings prefixed with F, red frame).
  \addtocounter{lstnofor}{1}%
  \renewcommand*\FancyVerbStartString{\PY{c}{!listing00}}
  \renewcommand*\FancyVerbStopString{\PY{c}{!listing01}}
  \setcounter{FancyVerbLine}{\thelinenofor}%
  \fvset{label={listing~F.\thelstnofor~(Fortran)},rulecolor=\color{red},stepnumber=1}%
  %
  \setcounter{linenofor}{\value{FancyVerbLine}}%

  Programming language constructs when inlined in the text are 
    typeset in bold, e.g. \prog{GOTO 2}.

  \section{Implementation}\label{sec:impl}

  Double precision floating-point format is used in all three implementations.
  The codes begin with the following definitions:
  \addtocounter{lstnocpp}{1}%
  \renewcommand*\FancyVerbStartString{\PY{c+c1}{//listing01}}%
  \renewcommand*\FancyVerbStopString{\PY{c+c1}{//listing02}}%
  \setcounter{FancyVerbLine}{\thelinenocpp}%
  \fvset{label={listing~C.\thelstnocpp~(C++)},rulecolor=\color{black},stepnumber=1}%
  \setcounter{linenocpp}{\value{FancyVerbLine}}%

  \addtocounter{lstnopyt}{1}%
  \renewcommand*\FancyVerbStartString{\PY{c}{\PYZsh{}listing01}}
  \renewcommand*\FancyVerbStopString{\PY{c}{\PYZsh{}listing02}}
  \setcounter{FancyVerbLine}{\thelinenopyt}%
  \fvset{label={listing~P.\thelstnopyt~(Python)},rulecolor=\color{blue},stepnumber=1}%
  \setcounter{linenopyt}{\value{FancyVerbLine}}%

  \addtocounter{lstnofor}{1}%
  \renewcommand*\FancyVerbStartString{\PY{c}{!listing01}}
  \renewcommand*\FancyVerbStopString{\PY{c}{!listing02}}
  \setcounter{FancyVerbLine}{\thelinenofor}%
  \fvset{label={listing~F.\thelstnofor~(Fortran)},rulecolor=\color{red},stepnumber=1}%
  \setcounter{linenofor}{\value{FancyVerbLine}}%

  which provide a convenient way of switching to different precision.

  All codes are structured in a way allowing compilation of the code 
    in exactly the same order as presented in the text within one source file,
    hence every Fortran listing contains definition of a separate module.

  \subsection{Array containers}\label{sec:array}

  Solution of equation (\ref{eq:adv}) using MPDATA implies discretisation onto a grid of the $\psi$ and 
    the Courant number $\vec{C}=\vec{v}\cdot\frac{\Delta t}{\Delta x}$ fields, where $\Delta t$ is the 
    solver timestep and $\Delta x$ is the grid spacing.

  Presented C++ implementation of MPDATA is built upon the Blitz++ 
    library\footnote{Blitz++ is a C++ class library for scientific computing which uses the expression templates technique to achieve high performance, see \url{http://sf.net/projects/blitz/}}.
  Blitz offers object-oriented representation of n-dimensional arrays,
    and array-valued mathematical expressions.
  In particular, it offers loop-free notation for array arithmetics
    that does not incur creation of intermediate temporary objects.
  Blitz++ is a header-only library\footnote{Blitz++ requires linking with \prog{libblitz} if debugging mode is used} -- to use it, it is enough to include the appropriate header file,
    and optionally expose the required classes to the present namespace:
  \addtocounter{lstnocpp}{1}%
  \renewcommand*\FancyVerbStartString{\PY{c+c1}{//listing02}}%
  \renewcommand*\FancyVerbStopString{\PY{c+c1}{//listing03}}%
  \setcounter{FancyVerbLine}{\thelinenocpp}%
  \fvset{label={listing~C.\thelstnocpp~(C++)},rulecolor=\color{black},stepnumber=1}%
  \setcounter{linenocpp}{\value{FancyVerbLine}}%

  Here \prog{arr\_t}, \prog{rng\_t} and \prog{idx\_t} serve as alias identifiers 
    and are introduced in order to shorten the code.

  The power of Blitz++ comes from the ability to express array expressions as objects.
  In particular, it is possible to define a function that returns an array expression;
    i.e. not the resultant array, but an object representing a ,,recipe'' defining the operations
    to be performed on the arguments.
  As a consequence, the return types of such functions become unintelligible.
  Luckily, the \prog{auto} return type declaration from the C++11 standard allows to simplify the code significantly,
    even more if used through the following preprocessor macro:
  \addtocounter{lstnocpp}{1}%
  \renewcommand*\FancyVerbStartString{\PY{c+c1}{//listing03}}%
  \renewcommand*\FancyVerbStopString{\PY{c+c1}{//listing04}}%
  \setcounter{FancyVerbLine}{\thelinenocpp}%
  \fvset{label={listing~C.\thelstnocpp~(C++)},rulecolor=\color{black},stepnumber=1}%
  \setcounter{linenocpp}{\value{FancyVerbLine}}%

  The call to \prog{blitz::safeToReturn()} function is included in order to ensure that
    all arrays involved in the expression being returned continue to exist in the
    caller scope.
  For example, definition of a function returning its array-valued argument doubled, reads:
    \prog{auto~f(arr\_t~x)~return\_macro(2*x)}.
  This is the only preprocessor macro defined herein.

  For the Python implementation of MPDATA the NumPy\footnote{NumPy is a Python package
    for scientific computing offering support for multi-dimensional arrays and a library
    of numerical algorithms, see \url{http://numpy.org/}} package is used.
  In order to make the code compatible with both the standard CPython
    as well as the alternative PyPy implementation of Python \citep[][]{Bolz_et_al_2011},
    the Python code includes the following sequence of \prog{import} statements:
  \addtocounter{lstnopyt}{1}%
  \renewcommand*\FancyVerbStartString{\PY{c}{\PYZsh{}listing02}}
  \renewcommand*\FancyVerbStopString{\PY{c}{\PYZsh{}listing03}}
  \setcounter{FancyVerbLine}{\thelinenopyt}%
  \fvset{label={listing~P.\thelstnopyt~(Python)},rulecolor=\color{blue},stepnumber=1}%
  \setcounter{linenopyt}{\value{FancyVerbLine}}%

  First, the PyPy's built-in NumPy implementation named \prog{numpypy} is imported if applicable (i.e. if running PyPy), 
    and the lazy evaluation mode is turned on through the \prog{set\_invalidation(False)} call.
  PyPy's lazy evaluation obtained with the help of a just-in-time compiler enables to achieve
    an analogous to Blitz++ temporary-array-free handling of array-valued expressions 
    (see discussion in section~\ref{sec:perf}).
  Second, to match the settings of C++ and Fortran compilers used herein, the NumPy package is instructed 
    to ignore any floating-point errors, if such an option
    is available in the interpreter\footnote{\prog{numpy.seterr()} is not supported in PyPy as of version 1.9}.
  The above lines conclude all code modifications that needed to be added in order to run
    the code with PyPy.

  Among the three considered languages only Fortran is equipped with built-in
    array handling facilities of practical use in high-performance computing. 
  Therefore, there is no need for using an external package as with C++ and Python.
  Fortran array-handling features are not object-oriented, though.

  \subsection{Containers for sequences of arrays}\label{sec:sequence}

  As discussed above, discretisation in space of the scalar field $\psi(x,y)$ into its $\psi_{[i,j]}$ 
    grid representation requires floating-point array containers.
  In turn, discretisation in time requires a container class for storing
    sequences of such arrays, i.e. \{$\psi^{[n]}$, $\psi^{[n+1]}$\}.
  Similarly the components of the vector field $\vec{C}$ are in fact a \{$C^{[x]}$, $C^{[y]}$\} 
    array sequence.
 
  Using an additional array dimension to represent the sequence elements is not considered for two reasons.
  First, the $C^{[x]}$ and $C^{[y]}$ arrays constituting the sequence have different sizes
    (see discussion of the Arakawa-C grid in section~\ref{sec:grid}).
  Second, the order of dimensions would need to be different for different languages to assure that
    the contiguous dimension is used for one of the space dimensions and not for time levels.

  In the C++ implementation the Boost\footnote{
    Boost is a free and open-source collection of peer-reviewed C++ libraries available at \url{http://www.boost.org/}.
    Several parts of Boost have been integrated into or inspired new additions to the C++ standard.
  } \prog{ptr\_vector} class is used to represent sequences of Blitz++ arrays 
    and at the same time to handle automatic freeing of dynamically allocated memory.
  The \prog{ptr\_vector} class is further customised by defining a derived structure which element-access \prog{[~]} 
    operator is overloaded with a modulo variant:
  \addtocounter{lstnocpp}{1}%
  \renewcommand*\FancyVerbStartString{\PY{c+c1}{//listing04}}%
  \renewcommand*\FancyVerbStopString{\PY{c+c1}{//listing05}}%
  \setcounter{FancyVerbLine}{\thelinenocpp}%
  \fvset{label={listing~C.\thelstnocpp~(C++)},rulecolor=\color{black},stepnumber=1}%
  \setcounter{linenocpp}{\value{FancyVerbLine}}%

  Consequently the last element of any such sequence may be accessed at index \prog{-1}, the last but one at \prog{-2}, 
    and so on.

  In the Python implementation the built-in \prog{tuple} type is used to store sequences of NumPy arrays.
  Employment of negative indices for handling from-the-end addressing of elements
    is a built-in feature of all sequence containers in Python.

  Fortran does not feature any built-in sequence container capable of storing arrays, 
    hence a custom \prog{arrvec\_t} type is introduced:
  \addtocounter{lstnofor}{1}%
  \renewcommand*\FancyVerbStartString{\PY{c}{!listing02}}
  \renewcommand*\FancyVerbStopString{\PY{c}{!listing03}}
  \setcounter{FancyVerbLine}{\thelinenofor}%
  \fvset{label={listing~F.\thelstnofor~(Fortran)},rulecolor=\color{red},stepnumber=1}%
  \setcounter{linenofor}{\value{FancyVerbLine}}%

  The \prog{arr\_t} type is defined solely for the purpose of overcoming the limitation 
    of lack of an array-of-arrays construct, and its only member field is a two-dimensional array.
  An array of \prog{arr\_t} is used hereinafter as a container for sequences of arrays.

  The \prog{arrptr\_t} type is defined solely for the purpose of overcoming Fortran's limitation
    of not supporting allocatables of pointers.
  \prog{arrptr\_t}'s single member field is a pointer to an instance of \prog{arr\_t}.
  Creating an allocatable of \prog{arrptr\_t}, instead of 
    a multi-element pointer of \prog{arr\_t}, ensures automatic memory deallocation.

  Type \prog{arrptr\_t} is used to implement the from-the-end addressing of elements in \prog{arrvec\_t}.
  The array data is stored in the \prog{arrs} member field (of type \prog{arr\_t}).
  The \prog{at} member field (of type \prog{arrptr\_t}) stores pointers to the elements of \prog{arrs}. 
  \prog{at} has double the length of \prog{arrs} and is initialised in a cyclic manner so that
    the \prog{-1} element of \prog{at} points to the last element of \prog{arrs}, and so on.
  Assuming \prog{psi} is an instance of \prog{arrptr\_t}, the \prog{(i,j)} element of the \prog{n}-th array in \prog{psi}
    may be accessed with\\ \prog{psi\%at( n )\%p\%a( i, j )}.

  The \prog{ctor(n)} method initialises the container for a given number of elements \prog{n}.
  The \prog{init(n,i,j)} method initialises the \prog{n}-th element of the container with 
    a newly allocated 2D array spanning indices \prog{i(1)}:\prog{i(2)}, 
    and \prog{j(1)}:\prog{j(2)} in the first, and last dimensions respectively\footnote{In Fortran, when an array
    is passed as a function argument its base is locally set to unity, regardless of the setting
    at the caller scope.}.

  \subsection{Staggered grid}\label{sec:grid}

  \begin{figure}[h!]
  \center
  \begin{tikzpicture}
    \coordinate (Origin)   at (0,0);
    \coordinate (XAxisMin) at (-2.5,0);
    \coordinate (XAxisMax) at (3,0);
    \coordinate (YAxisMin) at (0,-2.5);
    \coordinate (YAxisMax) at (0,3);
    \draw [thin, gray,-latex] (XAxisMin) -- (XAxisMax);
    \draw [thin, gray,-latex] (YAxisMin) -- (YAxisMax);

    \clip (-2.5,-2.5) rectangle (2.5cm,2.5cm); 
    \draw[style=help lines,dashed] (-14,-14) grid[step=2cm] (14,14);
    \foreach \x in {-7,-6,...,7}{
      \foreach \y in {-7,-6,...,7}{
        \node[draw,circle,inner sep=2pt,fill] at (2*\x,2*\y) {};
      }
    }
    \draw [black] (0,0) -- (0,0) node [below right] {$\!\psi_{[i,j]}$};
    \draw [black] (-2,0) -- (-2,0) node [below right] {$\!\psi_{[i-1,j]}$};
    \draw [black] (0,2) -- (0,2) node [below right] {$\!\psi_{[i,j+1]}$};

    \draw [ultra thick,-latex,red] (.8,0) -- (1.2,0) node [above right] {$\!\!\!\!\!\!\!\!\!\!C^{[x]}_{[i+\phlf,j]}$};
    \draw [ultra thick,-latex,red] (-1.2,0) -- (-.8,0) node [above right] {$\!\!\!\!\!\!\!\!\!\!C^{[x]}_{[i\mhlf,j]}$};
    \draw [ultra thick,-latex,red] (.8,2) -- (1.2,2) node [above] {};
    \draw [ultra thick,-latex,red] (-1.2,2) -- (-.8,2) node [above] {};
    \draw [ultra thick,-latex,red] (.8,-2) -- (1.2,-2) node [above] {};
    \draw [ultra thick,-latex,red] (-1.2,-2) -- (-.8,-2) node [above] {};

    \draw [ultra thick,-latex,red] (0,.8) -- (0,1.2) node [above] {};
    \draw [ultra thick,-latex,red] (0,-1.2) -- (0,-.8) node [below right] {$C^{[y]}_{[i,j\mhlf]}$};
    \draw [ultra thick,-latex,red] (-2,.8) -- (-2,1.2) node [above] {};
    \draw [ultra thick,-latex,red] (-2,-1.2) -- (-2,-.8) node [above] {};
    \draw [ultra thick,-latex,red] (2,.8) -- (2,1.2) node [above] {};
    \draw [ultra thick,-latex,red] (2,-1.2) -- (2,-.8) node [above] {};
  \end{tikzpicture}
  \caption{\label{fig:grid}
    A schematic of the Arakawa-C grid.
  }
  \end{figure}
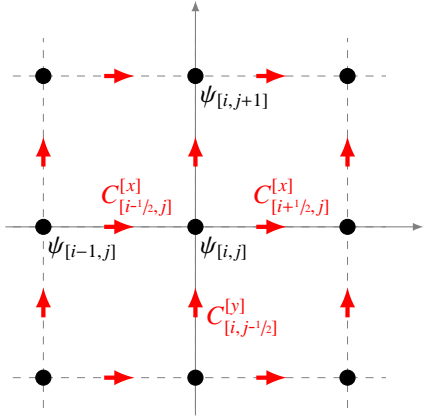
  The so-called Arakawa-C staggered grid \citep{Arakawa_and_Lamb_1977} depicted in Figure~\ref{fig:grid}
    is a natural choice for MPDATA.
  As a consequence, the discretised representations of the $\psi$ scalar field, and each 
    component of the ${\vec{C}=\vec{v}\cdot\frac{\Delta t}{\Delta x}}$ vector field 
    in eq.~(\ref{eq:adv}) are defined over different grid point locations.
  In mathematical notation this can be indicated by usage of fractional indices, e.g.
    $C^{[x]}_{[i\mhlf,j]}$, $C^{[x]}_{[i+\phlf,j]}$, $C^{[y]}_{[i,j\mhlf]}$ and $C^{[y]}_{[i,j+\phlf]}$
    to depict the grid values of the ${\vec{C}}$ vector components surrounding $\psi_{[i,j]}$.
  However, fractional indexing does not have a built-in counterpart in any of the
    employed programming languages.
  A desired syntax would translate \prog{$i-\phlf$} to \prog{$i-1$} and
    \prog{$i+\phlf$} to \prog{$i$}.
  OOP offers a convenient way to implement such notation
    by overloading the \prog{+} and \prog{-} operators for objects representing array indices. 

  In the C++ implementation first a global instance \prog{h} of an empty structure 
    \prog{hlf\_t} is defined, and then the plus and minus operators for \prog{hlf\_t} and \prog{rng\_t} are overloaded:
  \addtocounter{lstnocpp}{1}%
  \renewcommand*\FancyVerbStartString{\PY{c+c1}{//listing05}}%
  \renewcommand*\FancyVerbStopString{\PY{c+c1}{//listing06}}%
  \setcounter{FancyVerbLine}{\thelinenocpp}%
  \fvset{label={listing~C.\thelstnocpp~(C++)},rulecolor=\color{black},stepnumber=1}%
  \setcounter{linenocpp}{\value{FancyVerbLine}}%

  This way, the arrays representing vector field components can be indexed using
    \prog{(i+h,j)}, \prog{(i-h,j)} etc. where \prog{h}~represents the half.

  In NumPy in order to prevent copying of array data during slicing one needs to operate on the
    so-called array views.
  Array views are obtained when indexing the arrays with objects of the Python's
    built-it \prog{slice} type (or tuples of such objects in case of multi-dimensional arrays).
  Python forbids overloading of operators of built-in types such as \prog{slices}, 
    and does not define addition/subtraction operators for \prog{slice} and \prog{int} pairs.
  Consequently, a custom logic has to be defined not only for fractional indexing,
    but also for shifting the slices by integer intervals ($i\pm1$).
  It is implemented here by declaring a \prog{shift} class with the adequate operator overloads:
  \addtocounter{lstnopyt}{1}%
  \renewcommand*\FancyVerbStartString{\PY{c}{\PYZsh{}listing03}}
  \renewcommand*\FancyVerbStopString{\PY{c}{\PYZsh{}listing04}}
  \setcounter{FancyVerbLine}{\thelinenopyt}%
  \fvset{label={listing~P.\thelstnopyt~(Python)},rulecolor=\color{blue},stepnumber=1}%
  \setcounter{linenopyt}{\value{FancyVerbLine}}%

    and two instances of it to represent unity and half
    in expressions like \prog{i+one}, \prog{i+hlf}, where \prog{i} is an instance of \prog{slice}
    \footnote{\label{fnt:slice}One could argue that not using an own implementation of a slice-representing class
    in NumPy is a design flaw -- being able to modify behaviour of a hypothetical numpy.slice class 
    through inheritance would allow to implement the same behaviour as obtained in listing P.3 without the need to represent 
    the unity as a separate object}:
  \addtocounter{lstnopyt}{1}%
  \renewcommand*\FancyVerbStartString{\PY{c}{\PYZsh{}listing04}}
  \renewcommand*\FancyVerbStopString{\PY{c}{\PYZsh{}listing05}}
  \setcounter{FancyVerbLine}{\thelinenopyt}%
  \fvset{label={listing~P.\thelstnopyt~(Python)},rulecolor=\color{blue},stepnumber=1}%
  \setcounter{linenopyt}{\value{FancyVerbLine}}%

  In Fortran fractional array indexing is obtained through
    definition and instantiation of an object representing the half, and having appropriate
    operator overloads:
  \addtocounter{lstnofor}{1}%
  \renewcommand*\FancyVerbStartString{\PY{c}{!listing03}}
  \renewcommand*\FancyVerbStopString{\PY{c}{!listing04}}
  \setcounter{FancyVerbLine}{\thelinenofor}%
  \fvset{label={listing~F.\thelstnofor~(Fortran)},rulecolor=\color{red},stepnumber=1}%
  \setcounter{linenofor}{\value{FancyVerbLine}}%

  \subsection{Halo regions}

  The MPDATA formul\ae~defining $\psi^{[n+1]}_{[i,j]}$ as a function of $\psi^{[n]}_{[i,j]}$
    (discussed in the following sections) feature terms such as $\psi_{[i-1,j-1]}$.
  One way of assuring validity of these formul\ae~on the edges of the domain (e.g. for i=0) 
    is to introduce the so-called halo region surrounding the domain.
  The method of populating the halo region with data depends on the boundary condition type.
  Employment of the halo-region logic implies repeated usage of array range 
    extensions in the code such as $i \leadsto i \pm halo$.

  An \prog{ext()} function is defined in all three implementation, in order to simplify 
    coding of array range extensions:
  \addtocounter{lstnocpp}{1}%
  \renewcommand*\FancyVerbStartString{\PY{c+c1}{//listing06}}%
  \renewcommand*\FancyVerbStopString{\PY{c+c1}{//listing07}}%
  \setcounter{FancyVerbLine}{\thelinenocpp}%
  \fvset{label={listing~C.\thelstnocpp~(C++)},rulecolor=\color{black},stepnumber=1}%
  \setcounter{linenocpp}{\value{FancyVerbLine}}%

  \addtocounter{lstnopyt}{1}%
  \renewcommand*\FancyVerbStartString{\PY{c}{\PYZsh{}listing05}}
  \renewcommand*\FancyVerbStopString{\PY{c}{\PYZsh{}listing06}}
  \setcounter{FancyVerbLine}{\thelinenopyt}%
  \fvset{label={listing~P.\thelstnopyt~(Python)},rulecolor=\color{blue},stepnumber=1}%
  \setcounter{linenopyt}{\value{FancyVerbLine}}%

  \addtocounter{lstnofor}{1}%
  \renewcommand*\FancyVerbStartString{\PY{c}{!listing04}}
  \renewcommand*\FancyVerbStopString{\PY{c}{!listing05}}
  \setcounter{FancyVerbLine}{\thelinenofor}%
  \fvset{label={listing~F.\thelstnofor~(Fortran)},rulecolor=\color{red},stepnumber=1}%
  \setcounter{linenofor}{\value{FancyVerbLine}}%

  Consequently, a range depicted by $i\pm1/2$ may be expressed in the code as \prog{ext(i, h)}.
  In all three implementations the \prog{ext()} function accept the second
    argument to be an integer or a ''half'' (cf. section \ref{sec:grid}).

  \subsection{Array index permutations}\label{sec:pi}
  Hereinafter, the $\pi_{a,b}^{d}$ symbol is used to denote a  
    cyclic permutation of an order $d$ of a set $\{a,b\}$.
  It is used to generalise the MPDATA formul\ae~into multiple dimensions
    using the following notation:
  \begin{equation*}\label{eq:pi}
    \sum\limits_{d=0}^{1} \psi_{[i,j]+\pi_{\pone,\zero}^{d}} \equiv \psi_{[i+1,j]} + \psi_{[i,j+1]} 

  \end{equation*}

  Blitz++ ships with the \prog{RectDomain} class (aliased here as \prog{idx\_t}) 
    for specifying array ranges in multiple dimensions.
  The $\pi$ permutation is implemented in C++ as a function \prog{pi()}
    returning an instance of \prog{idx\_t}.
  In order to ensure compile-time evaluation, the permutation order is passed 
    via the template parameter \prog{d}
    (note the different order of \prog{i} and \prog{j} arguments in the two template specialisations):
  \addtocounter{lstnocpp}{1}%
  \renewcommand*\FancyVerbStartString{\PY{c+c1}{//listing07}}%
  \renewcommand*\FancyVerbStopString{\PY{c+c1}{//listing08}}%
  \setcounter{FancyVerbLine}{\thelinenocpp}%
  \fvset{label={listing~C.\thelstnocpp~(C++)},rulecolor=\color{black},stepnumber=1}%
  \setcounter{linenocpp}{\value{FancyVerbLine}}%

  NumPy uses tuples of slices for addressing multi-dimensional array with a
    single object.
  Therefore, the following definition of function \prog{pi()} suffices to represent $\pi$:
  \addtocounter{lstnopyt}{1}%
  \renewcommand*\FancyVerbStartString{\PY{c}{\PYZsh{}listing06}}
  \renewcommand*\FancyVerbStopString{\PY{c}{\PYZsh{}listing07}}
  \setcounter{FancyVerbLine}{\thelinenopyt}%
  \fvset{label={listing~P.\thelstnopyt~(Python)},rulecolor=\color{blue},stepnumber=1}%
  \setcounter{linenopyt}{\value{FancyVerbLine}}%

  In the Fortran implementation \prog{pi()} returns a pointer to the array elements specified
    by \prog{i} and \prog{j} interpreted as (i,j) or
    (j,i) depending on the value of the argument \prog{d}.
  In addition to \prog{pi()}, a helper \prog{span()} function returning the 
    length of one of the vectors passed as argument is defined:
  \addtocounter{lstnofor}{1}%
  \renewcommand*\FancyVerbStartString{\PY{c}{!listing05}}
  \renewcommand*\FancyVerbStopString{\PY{c}{!listing06}}
  \setcounter{FancyVerbLine}{\thelinenofor}%
  \fvset{label={listing~F.\thelstnofor~(Fortran)},rulecolor=\color{red},stepnumber=1}%
  \setcounter{linenofor}{\value{FancyVerbLine}}%

  The \prog{span()} function is used to shorten the declarations of arrays to be returned 
    from functions in the Fortran implementation (see listings F.11 and F.17--F.20).

  It is worth noting here that the C++ implementation of \prog{pi()} is
    branchless thanks to employment of template specialisation.
  With Fortran one needs to rely on compiler optimisations to eliminate the conditional expression
    within the \prog{pi()} that depends on value of \prog{d} which is always known 
    at compile time.

  \subsection{Prototype solver}\label{sec:solver}

  The tasks to be handled by a prototype advection equation solver proposed herein are:
  \begin{enumerate}[label=(\roman*), leftmargin=*, widest=iii]
    \item{storing arrays representing the $\psi$ and $\vec{C}$ fields and any required housekeeping data,}
    \item{allocating and deallocating the required memory,}
    \item{providing access to the solver state,}
    \item{performing the integration by invoking the advection-operator and
      boundary-condition handling routines.}
  \end{enumerate}
  In the following C++ definition of the \prog{solver} structure, task (i) is represented with the definition of the structure
    member fields; task (ii) is split between the \prog{solver}'s constructor and the destructors of \prog{arrvec\_t};
    task (iii) is handled by the accessor methods; task (iv) is handled within the \prog{solve} method:
  \addtocounter{lstnocpp}{1}%
  \renewcommand*\FancyVerbStartString{\PY{c+c1}{//listing08}}%
  \renewcommand*\FancyVerbStopString{\PY{c+c1}{//listing09}}%
  \setcounter{FancyVerbLine}{\thelinenocpp}%
  \fvset{label={listing~C.\thelstnocpp~(C++)},rulecolor=\color{black},stepnumber=1}%
  \setcounter{linenocpp}{\value{FancyVerbLine}}%

  The \prog{solver} structure is an abstract definition (containing a pure virtual method) 
     requiring its descendants to implement at least
     the \prog{advop()} method which is expected to fill \prog{psi[n+1]} with an updated (advected)
     values of \prog{psi[n]}.
  The two template parameters \prog{bcx\_t} and \prog{bcy\_t} allow the solver to operate with 
    any kind of boundary condition structures that fulfil the requirements implied by the calls to the 
    methods of \prog{bcx} and \prog{bcy}, respectively.

  The donor-cell and MPDATA schemes both require only the previous state of an advected field
    in order to advance the solution.
  Consequently, memory for two time levels ($\psi^{[n]}$ and $\psi^{[n+1]}$) is allocated in the
    constructor.
  The sizes of the arrays representing the two time levels of $\psi$ 
    are defined by the domain size ({\em nx}$~\times$~{\em ny}) plus the halo region.
  The size of the halo region is an argument of the constructor.
  The \prog{cycle()} method is used to swap the time levels without copying any data.

  The arrays representing the $C^{[x]}$ and $C^{[y]}$ components of $\vec{C}$,
    require ({\em nx+1})~$\times$~{\em ny} 
    and {\em nx}$~\times$~({\em ny+1}) elements, respectively
    (being laid out on the Arakawa-C staggered grid).
  
  Python definition of the \prog{solver} class follows closely the C++ structure definition:
  \addtocounter{lstnopyt}{1}%
  \renewcommand*\FancyVerbStartString{\PY{c}{\PYZsh{}listing07}}
  \renewcommand*\FancyVerbStopString{\PY{c}{\PYZsh{}listing08}}
  \setcounter{FancyVerbLine}{\thelinenopyt}%
  \fvset{label={listing~P.\thelstnopyt~(Python)},rulecolor=\color{blue},stepnumber=1}%
  \setcounter{linenopyt}{\value{FancyVerbLine}}%

  The key difference stems from the fact that, unlike Blitz++, NumPy does not allow an array
    to have arbitrary index base -- in NumPy the first element is always addressed with 0.
  Consequently, while in C++ (and Fortran) the computational domain is chosen to start at (i=0, j=0)
    and hence a part of the halo region to have negative indices, in Python the halo region starts 
    at (0,0)\footnote{The reason to allow the domain to begin at an arbitrary index
    is mainly to ease debugging in case the code would be used in parallel computations
    using domain decomposition where each subdomain could have its own index base corresponding to
    the location within the computational domain}.
  However, since the whole halo logic is hidden within the solver, such details are not exposed to the
    user.
  The \prog{bcx} and \prog{bcy} boundary-condition specifications are passed to the solver through
    constructor-like \prog{\_\_init\_\_()} method as opposed to template parameters in C++.

  The above C++ and Python prototype solvers in principle allow to operate with any boundary condition
    objects that implement methods called from within the solver.
  This requirement is checked at compile-time in the case of C++, and at run-time in the case of Python.
  In order to obtain an analogous behaviour with Fortran, it is required to
    define, prior to definition of a solver type, 
    an abstract type with deferred procedures having abstract interfaces 
    \citep[sic!, see Table 2.1 in][for a summary of approximate correspondence of OOP 
      nomenclature between Fortran and C++]{Rouson_et_al_2012}:
  \addtocounter{lstnofor}{1}%
  \renewcommand*\FancyVerbStartString{\PY{c}{!listing06}}
  \renewcommand*\FancyVerbStopString{\PY{c}{!listing07}}
  \setcounter{FancyVerbLine}{\thelinenofor}%
  \fvset{label={listing~F.\thelstnofor~(Fortran)},rulecolor=\color{red},stepnumber=1}%
  \setcounter{linenofor}{\value{FancyVerbLine}}%

  Having defined the abstract type for boundary-condition objects, 
    a definition of a solver class following closely the C++ and Python counterparts may be provided:
  \addtocounter{lstnofor}{1}%
  \renewcommand*\FancyVerbStartString{\PY{c}{!listing07}}
  \renewcommand*\FancyVerbStopString{\PY{c}{!listing08}}
  \setcounter{FancyVerbLine}{\thelinenofor}%
  \fvset{label={listing~F.\thelstnofor~(Fortran)},rulecolor=\color{red},stepnumber=1}%
  \setcounter{linenofor}{\value{FancyVerbLine}}%

  \subsection{Periodic boundaries (C++)}\label{sec:cyclic}

  From this point, only C++ implementation is explained in the main text. 
  The Python and Fortran implementations are included in appendices P and F.  

  The solver definition described in section~\ref{sec:solver} requires a given
    boundary condition object to implement a \prog{fill\_halos()} method.
  An implementation of periodic boundary conditions in C++ is provided in the following listing:
  \addtocounter{lstnocpp}{1}%
  \renewcommand*\FancyVerbStartString{\PY{c+c1}{//listing09}}%
  \renewcommand*\FancyVerbStopString{\PY{c+c1}{//listing10}}%
  \setcounter{FancyVerbLine}{\thelinenocpp}%
  \fvset{label={listing~C.\thelstnocpp~(C++)},rulecolor=\color{black},stepnumber=1}%
  \setcounter{linenocpp}{\value{FancyVerbLine}}%

  As hinted by the member field names, the \prog{fill\_halos()} methods
    fill the left/right halo regions with data from the right/left edges of the domain.
  Thanks to employment of the function \prog{pi()} described in section~\ref{sec:pi}
    the same code may be applied in any dimension (here being a template parameter).

  Listings P.8 and F.8 contain the Python and Fortran counterparts to listing C.9.

  \subsection{Donor-cell formul\ae~(C++)}\label{sec:donor}
  MPDATA is an iterative algorithm in which each iteration takes the form of the 
    so-called donor-cell formula (which itself is a first-order advection scheme).

  MPDATA and donor-cell are explicit forward-in-time algorithms -- they allow to predict $\psi^{[n+1]}$ as a
    function of $\psi^{[n]}$ where $n$ and $n+1$ denote two adjacent time levels.
  The donor-cell scheme may be written as \citep[eq. 2 in][]{Smolarkiewicz_1984}:
  \begin{equation}\label{eq:donor}
      \begin{split}
    \psi_{[i,j]}^{[n+1]} \!= \psi_{[i,j]}^{[n]}\!-\!\!\sum\limits_{d=0}^{N-1}\!\left(\!
      F\!\left[\psi_{[i,j]}^{[n]}, \psi^{[n]}_{[i,j]+\pi_{\pone,\zero}^{d}}\!, C^{[d]}_{[i,j]+\pi_{\phlf,\zero}^{d}}\right]
      \right.\\
      \left.-
      F\!\left[\psi_{[i,j]+\pi_{\mone,\zero}^{d}}^{[n]}\!, \psi^{[n]}_{[i,j]}, C^{[d]}_{[i,j]+\pi_{\mhlf,\zero}^{d}}\right]
    \right)\!\!\!\!
  \end{split}

  \end{equation}
  where $N$ is the number of dimensions, 
    and F is the so-called flux function \citep[][eq.~3]{Smolarkiewicz_1984}:
  \begin{equation}\label{eq:donor:F}
        \begin{split}
    F(\psi_L, \psi_R, C) = {\rm max}(C,0) \cdot \psi_L + {\rm min}(C,0) \cdot \psi_R\\
                         = \,\frac{C + |C|}{2}\,\, \cdot \psi_L + \,\frac{C - |C|}{2}\,\, \cdot \psi_R
    \end{split}

  \end{equation}

  \noindent The flux function takes the following form in C++:
  \addtocounter{lstnocpp}{1}%
  \renewcommand*\FancyVerbStartString{\PY{c+c1}{//listing10}}%
  \renewcommand*\FancyVerbStopString{\PY{c+c1}{//listing11}}%
  \setcounter{FancyVerbLine}{\thelinenocpp}%
  \fvset{label={listing~C.\thelstnocpp~(C++)},rulecolor=\color{black},stepnumber=1}%
  \setcounter{linenocpp}{\value{FancyVerbLine}}%

  \noindent 
  Equation~\ref{eq:donor} is split into the terms under the summation 
    (effectively the 1-dimensional donor-cell formula):
  \addtocounter{lstnocpp}{1}%
  \renewcommand*\FancyVerbStartString{\PY{c+c1}{//listing11}}%
  \renewcommand*\FancyVerbStopString{\PY{c+c1}{//listing12}}%
  \setcounter{FancyVerbLine}{\thelinenocpp}%
  \fvset{label={listing~C.\thelstnocpp~(C++)},rulecolor=\color{black},stepnumber=1}%
  \setcounter{linenocpp}{\value{FancyVerbLine}}%

  \noindent and the actual two-dimensional donor-cell formula:

  \addtocounter{lstnocpp}{1}%
  \renewcommand*\FancyVerbStartString{\PY{c+c1}{//listing12}}%
  \renewcommand*\FancyVerbStopString{\PY{c+c1}{//listing13}}%
  \setcounter{FancyVerbLine}{\thelinenocpp}%
  \fvset{label={listing~C.\thelstnocpp~(C++)},rulecolor=\color{black},stepnumber=1}%
  \setcounter{linenocpp}{\value{FancyVerbLine}}%

  Listings P.9-P11 and F.9-F.13 contain the Python and Fortran counterparts to listings C.12-C.15.

  \subsection{Donor-cell solver (C++)}\label{sec:donorcell_solver}
  
  As mentioned in the previous section, the donor-cell formula
    constitutes an advection scheme, hence we may use it to create
    a \prog{solver\_donorcell} implementation of the abstract \prog{solver} class: 
  \addtocounter{lstnocpp}{1}%
  \renewcommand*\FancyVerbStartString{\PY{c+c1}{//listing13}}%
  \renewcommand*\FancyVerbStopString{\PY{c+c1}{//listing14}}%
  \setcounter{FancyVerbLine}{\thelinenocpp}%
  \fvset{label={listing~C.\thelstnocpp~(C++)},rulecolor=\color{black},stepnumber=1}%
  \setcounter{linenocpp}{\value{FancyVerbLine}}%

  The above definition is given as an example only.
  In the following sections an MPDATA solver of the same structure is defined.

  Listings P.12 and F.14 contain the Python and Fortran counterparts to listing C.16.

  \subsection{MPDATA formul\ae~(C++)}\label{sec:mpdata}

  MPDATA introduces corrective steps to the algorithm defined by equation~\ref{eq:donor} and \ref{eq:donor:F}.
  Each corrective step is a donor-cell step (eq.~\ref{eq:donor}) with the Courant number
    fields corresponding to the MPDATA antidiffusive velocities of the following 
    form \citep[eqs~13,~14~in][]{Smolarkiewicz_1984}:
  \begin{equation}\label{eq:antidiff1}
        \begin{split}
      C'^{[d]}_{[i,j]+\pi^d_{\phlf,\zero}} \!\!\!=\! 
        \left| C^{[d]}_{[i,j] + \pi^d_{\phlf,\zero}} \right| 
        \!\cdot\! \left[ 1 - \left| C^{[d]}_{[i,j]+\pi_{\phlf,\zero}^d} \right| \right] 
        \!\!\cdot\! A^{[d]}_{[i,j]}(\psi) \\
        - \sum\limits_{q=0, q \ne d}^{N} C^{[d]}_{[i,j]+\pi_{\phlf,\zero}^d} 
        \!\!\cdot \overline{C}^{[q]}_{[i,j]+\pi_{\phlf,\zero}^d} \!\!\cdot B^{[d]}_{[i,j]}(\psi)
    \end{split}

  \end{equation}
  where $\psi$ and $C$ represent values from the previous iteration and where:
  \begin{equation}\label{eq:antidiff2}
        \begin{split}
      \overline{C}^{[q]}_{[i,j]+\pi^d_{\phlf,\zero}} \!\!= \frac{1}{4} \cdot \left(
          C^{[q]}_{[i,j]+\pi^d_{\pone,\phlf}} \!\!+ 
          C^{[q]}_{[i,j]+\pi^d_{\zero,\phlf}} 
        \right.+\\ \left.
          C^{[q]}_{[i,j]+\pi^d_{\pone,\mhlf}} \!\!+ 
          C^{[q]}_{[i,j]+\pi^d_{\zero,\mhlf}}
      \right)
    \end{split}

  \end{equation}
  For positive-definite $\psi$, the $A$ and $B$ terms take the 
  following form\footnote{
    Since $\psi\ge0$, $|A|\le1$ and $|B|\le1$. 
    See \citet[Sec. 4.2]{Smolarkiewicz_2006} for description of adaptation of the 
    formul\ae~for advection of fields of variable sign
  }:
  \begin{equation}\label{eq:A}
        A^{[d]}_{[i,j]} \!= 
      \frac{
        \psi_{[i,j]+\pi^d_{\pone,\zero}} \!- \psi_{[i,j]}
      }{
        \psi_{[i,j]+\pi^d_{\pone,\zero}} \!+ \psi_{[i,j]}
      }

  \end{equation}
  \begin{equation}\label{eq:B}
        B^{[d]}_{[i,j]} \!= \frac{1}{2}\frac{
      \psi_{[i,j]+\pi^d_{\pone,\pone}} \!\!+ 
      \psi_{[i,j]+\pi^d_{\zero,\pone}} \!\!- 
      \psi_{[i,j]+\pi^d_{\pone,\mone}} \!\!- 
      \psi_{[i,j]+\pi^d_{\zero,\mone}}
    }{
      \psi_{[i,j]+\pi^d_{\pone,\pone}} \!\!+ 
      \psi_{[i,j]+\pi^d_{\zero,\pone}} \!\!+ 
      \psi_{[i,j]+\pi^d_{\pone,\mone}} \!\!+ 
      \psi_{[i,j]+\pi^d_{\zero,\mone}}
    }

  \end{equation}

  If the denominator in equations \ref{eq:A} or \ref{eq:B} equals zero for a given {\em i} and {\em j}, 
    the corresponding $A_{[i,j]}$ and $B_{[i,j]}$ are set to zero what may be conveniently 
    represented with the \prog{where} construct (available in all three considered languages):
  \addtocounter{lstnocpp}{1}%
  \renewcommand*\FancyVerbStartString{\PY{c+c1}{//listing14}}%
  \renewcommand*\FancyVerbStopString{\PY{c+c1}{//listing15}}%
  \setcounter{FancyVerbLine}{\thelinenocpp}%
  \fvset{label={listing~C.\thelstnocpp~(C++)},rulecolor=\color{black},stepnumber=1}%
  \setcounter{linenocpp}{\value{FancyVerbLine}}%

  The $A$ term defined in equation \ref{eq:A} takes the following form:
  \addtocounter{lstnocpp}{1}%
  \renewcommand*\FancyVerbStartString{\PY{c+c1}{//listing15}}%
  \renewcommand*\FancyVerbStopString{\PY{c+c1}{//listing16}}%
  \setcounter{FancyVerbLine}{\thelinenocpp}%
  \fvset{label={listing~C.\thelstnocpp~(C++)},rulecolor=\color{black},stepnumber=1}%
  \setcounter{linenocpp}{\value{FancyVerbLine}}%

  The $B$ term defined in equation \ref{eq:B} takes the following form:
  \addtocounter{lstnocpp}{1}%
  \renewcommand*\FancyVerbStartString{\PY{c+c1}{//listing16}}%
  \renewcommand*\FancyVerbStopString{\PY{c+c1}{//listing17}}%
  \setcounter{FancyVerbLine}{\thelinenocpp}%
  \fvset{label={listing~C.\thelstnocpp~(C++)},rulecolor=\color{black},stepnumber=1}%
  \setcounter{linenocpp}{\value{FancyVerbLine}}%

  Equation \ref{eq:antidiff2} takes the following form:
  \addtocounter{lstnocpp}{1}%
  \renewcommand*\FancyVerbStartString{\PY{c+c1}{//listing17}}%
  \renewcommand*\FancyVerbStopString{\PY{c+c1}{//listing18}}%
  \setcounter{FancyVerbLine}{\thelinenocpp}%
  \fvset{label={listing~C.\thelstnocpp~(C++)},rulecolor=\color{black},stepnumber=1}%
  \setcounter{linenocpp}{\value{FancyVerbLine}}%

  Equation \ref{eq:antidiff1} take the following form:
  \addtocounter{lstnocpp}{1}%
  \renewcommand*\FancyVerbStartString{\PY{c+c1}{//listing18}}%
  \renewcommand*\FancyVerbStopString{\PY{c+c1}{//listing19}}%
  \setcounter{FancyVerbLine}{\thelinenocpp}%
  \fvset{label={listing~C.\thelstnocpp~(C++)},rulecolor=\color{black},stepnumber=1}%
  \setcounter{linenocpp}{\value{FancyVerbLine}}%

  Listings P.13-P.17 and F.15-F.21 contain the Python and Fortran counterparts to listing C.16-C.22.

  \subsection{MPDATA solver (C++)}\label{sec:mpdatasolver}

  An MPDATA solver may be now constructed by inheriting from \prog{solver} class
    with the following definition in C++:
  \addtocounter{lstnocpp}{1}%
  \renewcommand*\FancyVerbStartString{\PY{c+c1}{//listing19}}%
  \renewcommand*\FancyVerbStopString{\PY{c+c1}{//listing20}}%
  \setcounter{FancyVerbLine}{\thelinenocpp}%
  \fvset{label={listing~C.\thelstnocpp~(C++)},rulecolor=\color{black},stepnumber=1}%
  \setcounter{linenocpp}{\value{FancyVerbLine}}%

  The array of sequences of temporary arrays \prog{tmp} allocated in the constructor
    is used to store the antidiffusive velocities from the present and optionally 
    previous timestep (if using more than two iterations). 

  The \prog{advop()} method controlls the MPDATA iterations within one timestep.
  The first (step = 0) iteration of MPDATA is an unmodified donor-cell step (compare listing C.15).
  Subsequent iterations begin with calculation of the antidiffusive Courant fields using 
    formula \ref{eq:antidiff1}.
  In order to calculate values spanning an (i\mhlf~... i+\phlf) range using a formula
    for $C_{[i+\phlf,\ldots]}$ only, the formula is evaluated using extended
    index ranges \prog{im} and \prog{jm}.
  In the second (step=1) iteration the uncorrected Courant field (\prog{C\_unco}) points
    to the original \prog{C} field, and the antidiffusive Courant field is written into
    \prog{C\_corr} which points to \prog{tmp[1]}.
  In the third (step=2) iteration \prog{C\_unco} points to \prog{tmp[1]} while \prog{C\_corr}
    points to \prog{tmp[0]}.
  In subsequent iterations \prog{tmp[0]} and \prog{tmp[1]} are alternately swapped.

  Listings P.18 and F.22 contain the Python and Fortran counterparts to listing C.23.
 
  \subsection{Usage example (C++)}\label{sec:example}

  The following listing provides an example of how the MPDATA solver
    defined in section~\ref{sec:mpdatasolver} may be used together
    with the cyclic boundary conditions defined in section~\ref{sec:cyclic}.
  In the example a Gaussian signal is advected in a 2D domain 
    defined over a grid of 24$\times$24 cells.
  The program first plots the initial condition, then performs
    the integration for 75 timesteps with three different 
    settings of the number of iterations used in MPDATA.
  The velocity field is constant in time and space (although it is
    not assumed in the presented implementations).
  The signal shape at the end of each simulation is plotted as well.
  Plotting is done with the help of the gnuplot-iostream library\footnote{gnuplot-iostream
    is a header-only C++ library allowing gnuplot to be controlled from C++, see \url{http://stahlke.org/dan/gnuplot-iostream/}. 
    Gnuplot is a portable command-line driven graphing utility, see \url{http://gnuplot.info/}}.

\begin{figure}
    \includegraphics[height=.9\textheight]{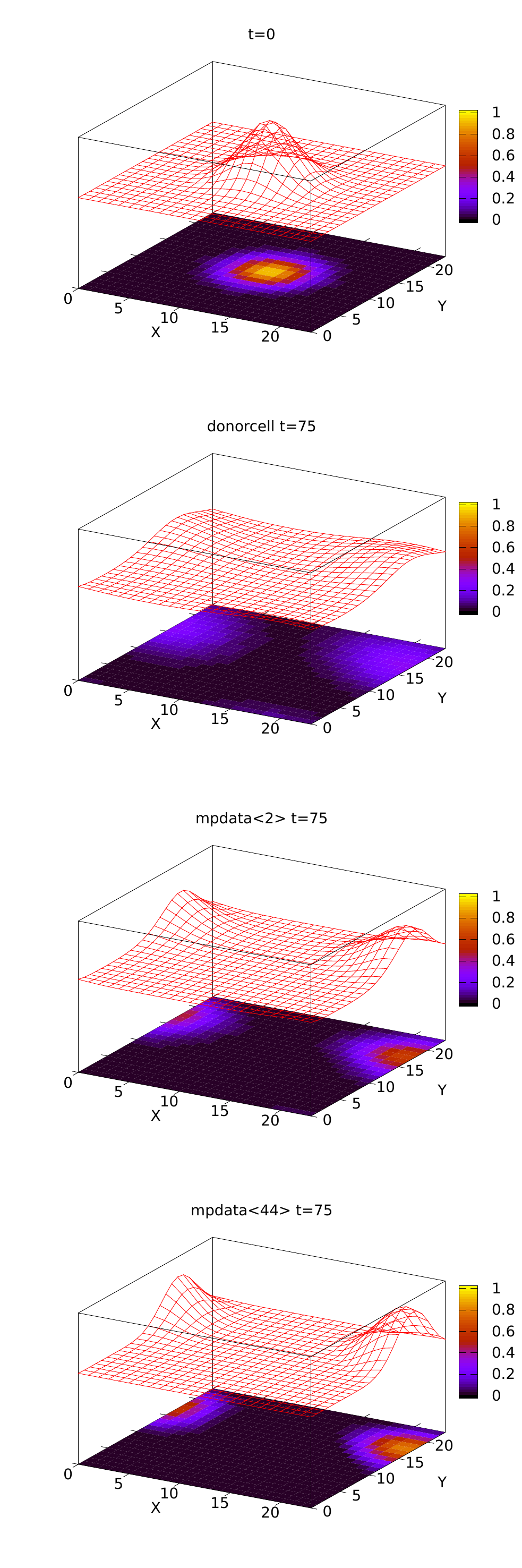}
    \caption{\label{fig:mpdata}
      Plot generated by the program given in listing~C.24.
      The top panel shows initial signal shape (at time t=0).
      The subsequent panels show snapshots of the advected field after 75 timesteps
        from three different simulations: donorcell (or 1 MPDATA iteration), MPDATA
        with two iterations and MPDATA with 44 iterations.
      The colour scale and the wire-frame surface correspond to signal amplitude.
      See section~\ref{sec:example} for discussion.
    }
  \end{figure}

  The resultant plot is presented herein as Figure~\ref{fig:mpdata}.
  The top panel depicts the initial condition. 
  The three other panels show a snapshot of the field after 75 timesteps.
  The donor-cell solution is characterised by strongest numerical diffusion
    resulting in significant drop in the signal amplitude.
  The signals advected using MPDATA show smaller numerical diffusion with
    the solution obtained with more iterations preserving the signal altitude 
    more accurately.
  In all of the simulations the signal maintains its positive definiteness.
  The domain periodicity is apparent in the plots as the maximum of the signal 
    after 75 timesteps is located near the domain walls.

  Listings P.19 and F.23-F.24 contain the Python and Fortran 
    counterparts to listing C.24 (with the set-up and plotting logic omitted).

  \addtocounter{lstnocpp}{1}%
  \renewcommand*\FancyVerbStartString{\PY{c+c1}{//listing20}}%
  \renewcommand*\FancyVerbStopString{\PY{c+c1}{//listing21}}%
  \setcounter{FancyVerbLine}{\thelinenocpp}%
  \fvset{label={listing~C.\thelstnocpp~(C++)},rulecolor=\color{black},stepnumber=1}%
  \begin{Verbatim}[commandchars=\\\{\}]
\PY{c+c1}{//listing20}
\PY{c+cp}{\PYZsh{}}\PY{c+cp}{include "listings.hpp"}
\PY{c+cp}{\PYZsh{}}\PY{c+cp}{define GNUPLOT\PYZus{}ENABLE\PYZus{}BLITZ}
\PY{c+cp}{\PYZsh{}}\PY{c+cp}{include \PYZlt{}gnuplot-iostream}\PY{c+cp}{/}\PY{c+cp}{gnuplot-iostream.h\PYZgt{}}

\PY{k}{enum} \PY{p}{\PYZob{}}\PY{n}{x}\PY{p}{,} \PY{n}{y}\PY{p}{\PYZcb{}}\PY{p}{;}

\PY{k}{template} \PY{o}{\PYZlt{}}\PY{k}{class} \PY{n+nc}{T}\PY{o}{\PYZgt{}}
\PY{k+kt}{void} \PY{n}{setup}\PY{p}{(}\PY{n}{T} \PY{o}{\PYZam{}}\PY{n}{solver}\PY{p}{,} \PY{k+kt}{int} \PY{n}{n}\PY{p}{[}\PY{l+m+mi}{2}\PY{p}{]}\PY{p}{)} 
\PY{p}{\PYZob{}}
  \PY{n}{blitz}\PY{o}{:}\PY{o}{:}\PY{n}{firstIndex} \PY{n}{i}\PY{p}{;}
  \PY{n}{blitz}\PY{o}{:}\PY{o}{:}\PY{n}{secondIndex} \PY{n}{j}\PY{p}{;}
  \PY{n}{solver}\PY{p}{.}\PY{n}{state}\PY{p}{(}\PY{p}{)} \PY{o}{=} \PY{n}{exp}\PY{p}{(}
    \PY{o}{-}\PY{n}{sqr}\PY{p}{(}\PY{n}{i}\PY{o}{-}\PY{n}{n}\PY{p}{[}\PY{n}{x}\PY{p}{]}\PY{o}{/}\PY{l+m+mf}{2.}\PY{p}{)} \PY{o}{/} \PY{p}{(}\PY{l+m+mi}{2}\PY{o}{*}\PY{n}{pow}\PY{p}{(}\PY{n}{n}\PY{p}{[}\PY{n}{x}\PY{p}{]}\PY{o}{/}\PY{l+m+mf}{10.}\PY{p}{,} \PY{l+m+mi}{2}\PY{p}{)}\PY{p}{)}
    \PY{o}{-}\PY{n}{sqr}\PY{p}{(}\PY{n}{j}\PY{o}{-}\PY{n}{n}\PY{p}{[}\PY{n}{y}\PY{p}{]}\PY{o}{/}\PY{l+m+mf}{2.}\PY{p}{)} \PY{o}{/} \PY{p}{(}\PY{l+m+mi}{2}\PY{o}{*}\PY{n}{pow}\PY{p}{(}\PY{n}{n}\PY{p}{[}\PY{n}{y}\PY{p}{]}\PY{o}{/}\PY{l+m+mf}{10.}\PY{p}{,} \PY{l+m+mi}{2}\PY{p}{)}\PY{p}{)}
  \PY{p}{)}\PY{p}{;}  
  \PY{n}{solver}\PY{p}{.}\PY{n}{courant}\PY{p}{(}\PY{n}{x}\PY{p}{)} \PY{o}{=} \PY{o}{-}\PY{l+m+mf}{.5}\PY{p}{;} 
  \PY{n}{solver}\PY{p}{.}\PY{n}{courant}\PY{p}{(}\PY{n}{y}\PY{p}{)} \PY{o}{=} \PY{o}{-}\PY{l+m+mf}{.25}\PY{p}{;}
\PY{p}{\PYZcb{}}

\PY{k+kt}{int} \PY{n}{main}\PY{p}{(}\PY{p}{)} 
\PY{p}{\PYZob{}}
  \PY{k+kt}{int} \PY{n}{n}\PY{p}{[}\PY{p}{]} \PY{o}{=} \PY{p}{\PYZob{}}\PY{l+m+mi}{24}\PY{p}{,} \PY{l+m+mi}{24}\PY{p}{\PYZcb{}}\PY{p}{,} \PY{n}{nt} \PY{o}{=} \PY{l+m+mi}{75}\PY{p}{;}
  \PY{n}{Gnuplot} \PY{n}{gp}\PY{p}{;}
  \PY{n}{gp} \PY{o}{\PYZlt{}}\PY{o}{\PYZlt{}} \PY{l+s}{"}\PY{l+s}{set term pdf size 10cm, 30cm}\PY{l+s+se}{\PYZbs{}n}\PY{l+s}{"} 
     \PY{o}{\PYZlt{}}\PY{o}{\PYZlt{}} \PY{l+s}{"}\PY{l+s}{set output 'figure.pdf'}\PY{l+s+se}{\PYZbs{}n}\PY{l+s}{"}     
     \PY{o}{\PYZlt{}}\PY{o}{\PYZlt{}} \PY{l+s}{"}\PY{l+s}{set multiplot layout 4,1}\PY{l+s+se}{\PYZbs{}n}\PY{l+s}{"} 
     \PY{o}{\PYZlt{}}\PY{o}{\PYZlt{}} \PY{l+s}{"}\PY{l+s}{set border 4095}\PY{l+s+se}{\PYZbs{}n}\PY{l+s}{"}
     \PY{o}{\PYZlt{}}\PY{o}{\PYZlt{}} \PY{l+s}{"}\PY{l+s}{set xtics out}\PY{l+s+se}{\PYZbs{}n}\PY{l+s}{"}
     \PY{o}{\PYZlt{}}\PY{o}{\PYZlt{}} \PY{l+s}{"}\PY{l+s}{set ytics out}\PY{l+s+se}{\PYZbs{}n}\PY{l+s}{"}
     \PY{o}{\PYZlt{}}\PY{o}{\PYZlt{}} \PY{l+s}{"}\PY{l+s}{unset ztics}\PY{l+s+se}{\PYZbs{}n}\PY{l+s}{"}    
     \PY{o}{\PYZlt{}}\PY{o}{\PYZlt{}} \PY{l+s}{"}\PY{l+s}{set xlabel 'X'}\PY{l+s+se}{\PYZbs{}n}\PY{l+s}{"}
     \PY{o}{\PYZlt{}}\PY{o}{\PYZlt{}} \PY{l+s}{"}\PY{l+s}{set ylabel 'Y'}\PY{l+s+se}{\PYZbs{}n}\PY{l+s}{"}
     \PY{o}{\PYZlt{}}\PY{o}{\PYZlt{}} \PY{l+s}{"}\PY{l+s}{set xrange [0:}\PY{l+s}{"} \PY{o}{\PYZlt{}}\PY{o}{\PYZlt{}} \PY{n}{n}\PY{p}{[}\PY{n}{x}\PY{p}{]}\PY{o}{-}\PY{l+m+mi}{1} \PY{o}{\PYZlt{}}\PY{o}{\PYZlt{}} \PY{l+s}{"}\PY{l+s}{]}\PY{l+s+se}{\PYZbs{}n}\PY{l+s}{"}   
     \PY{o}{\PYZlt{}}\PY{o}{\PYZlt{}} \PY{l+s}{"}\PY{l+s}{set yrange [0:}\PY{l+s}{"} \PY{o}{\PYZlt{}}\PY{o}{\PYZlt{}} \PY{n}{n}\PY{p}{[}\PY{n}{y}\PY{p}{]}\PY{o}{-}\PY{l+m+mi}{1} \PY{o}{\PYZlt{}}\PY{o}{\PYZlt{}} \PY{l+s}{"}\PY{l+s}{]}\PY{l+s+se}{\PYZbs{}n}\PY{l+s}{"}   
     \PY{o}{\PYZlt{}}\PY{o}{\PYZlt{}} \PY{l+s}{"}\PY{l+s}{set zrange [-.666:1]}\PY{l+s+se}{\PYZbs{}n}\PY{l+s}{"}   
     \PY{o}{\PYZlt{}}\PY{o}{\PYZlt{}} \PY{l+s}{"}\PY{l+s}{set cbrange [-.025:1.025]}\PY{l+s+se}{\PYZbs{}n}\PY{l+s}{"}     
     \PY{o}{\PYZlt{}}\PY{o}{\PYZlt{}} \PY{l+s}{"}\PY{l+s}{set palette maxcolors 42}\PY{l+s+se}{\PYZbs{}n}\PY{l+s}{"}
     \PY{o}{\PYZlt{}}\PY{o}{\PYZlt{}} \PY{l+s}{"}\PY{l+s}{set pm3d at b}\PY{l+s+se}{\PYZbs{}n}\PY{l+s}{"}\PY{p}{;}
  \PY{n}{std}\PY{o}{:}\PY{o}{:}\PY{n}{string} \PY{n}{binfmt}\PY{p}{;}
  \PY{p}{\PYZob{}}
    \PY{n}{solver\PYZus{}donorcell}\PY{o}{\PYZlt{}}\PY{n}{cyclic}\PY{o}{\PYZlt{}}\PY{n}{x}\PY{o}{\PYZgt{}}\PY{p}{,} \PY{n}{cyclic}\PY{o}{\PYZlt{}}\PY{n}{y}\PY{o}{\PYZgt{}}\PY{o}{\PYZgt{}} 
      \PY{n}{slv}\PY{p}{(}\PY{n}{n}\PY{p}{[}\PY{n}{x}\PY{p}{]}\PY{p}{,} \PY{n}{n}\PY{p}{[}\PY{n}{y}\PY{p}{]}\PY{p}{)}\PY{p}{;}
    \PY{n}{setup}\PY{p}{(}\PY{n}{slv}\PY{p}{,} \PY{n}{n}\PY{p}{)}\PY{p}{;}
    \PY{n}{binfmt} \PY{o}{=} \PY{n}{gp}\PY{p}{.}\PY{n}{binfmt}\PY{p}{(}\PY{n}{slv}\PY{p}{.}\PY{n}{state}\PY{p}{(}\PY{p}{)}\PY{p}{)}\PY{p}{;}
    \PY{n}{gp} \PY{o}{\PYZlt{}}\PY{o}{\PYZlt{}} \PY{l+s}{"}\PY{l+s}{set title 't=0'}\PY{l+s+se}{\PYZbs{}n}\PY{l+s}{"}
       \PY{o}{\PYZlt{}}\PY{o}{\PYZlt{}} \PY{l+s}{"}\PY{l+s}{splot '-' binary}\PY{l+s}{"} \PY{o}{\PYZlt{}}\PY{o}{\PYZlt{}} \PY{n}{binfmt}
       \PY{o}{\PYZlt{}}\PY{o}{\PYZlt{}} \PY{l+s}{"}\PY{l+s}{with lines notitle}\PY{l+s+se}{\PYZbs{}n}\PY{l+s}{"}\PY{p}{;}
    \PY{n}{gp}\PY{p}{.}\PY{n}{sendBinary}\PY{p}{(}\PY{n}{slv}\PY{p}{.}\PY{n}{state}\PY{p}{(}\PY{p}{)}\PY{p}{.}\PY{n}{copy}\PY{p}{(}\PY{p}{)}\PY{p}{)}\PY{p}{;}
    \PY{n}{slv}\PY{p}{.}\PY{n}{solve}\PY{p}{(}\PY{n}{nt}\PY{p}{)}\PY{p}{;}
    \PY{n}{gp} \PY{o}{\PYZlt{}}\PY{o}{\PYZlt{}} \PY{l+s}{"}\PY{l+s}{set title 'donorcell t=}\PY{l+s}{"}\PY{o}{\PYZlt{}}\PY{o}{\PYZlt{}}\PY{n}{nt}\PY{o}{\PYZlt{}}\PY{o}{\PYZlt{}}\PY{l+s}{"}\PY{l+s}{'}\PY{l+s+se}{\PYZbs{}n}\PY{l+s}{"}
       \PY{o}{\PYZlt{}}\PY{o}{\PYZlt{}} \PY{l+s}{"}\PY{l+s}{splot '-' binary}\PY{l+s}{"} \PY{o}{\PYZlt{}}\PY{o}{\PYZlt{}} \PY{n}{binfmt}
       \PY{o}{\PYZlt{}}\PY{o}{\PYZlt{}} \PY{l+s}{"}\PY{l+s}{with lines notitle}\PY{l+s+se}{\PYZbs{}n}\PY{l+s}{"}\PY{p}{;}
    \PY{n}{gp}\PY{p}{.}\PY{n}{sendBinary}\PY{p}{(}\PY{n}{slv}\PY{p}{.}\PY{n}{state}\PY{p}{(}\PY{p}{)}\PY{p}{.}\PY{n}{copy}\PY{p}{(}\PY{p}{)}\PY{p}{)}\PY{p}{;}
  \PY{p}{\PYZcb{}} 
  \PY{p}{\PYZob{}}
    \PY{k}{const} \PY{k+kt}{int} \PY{n}{it} \PY{o}{=} \PY{l+m+mi}{2}\PY{p}{;}
    \PY{n}{solver\PYZus{}mpdata}\PY{o}{\PYZlt{}}\PY{n}{it}\PY{p}{,} \PY{n}{cyclic}\PY{o}{\PYZlt{}}\PY{n}{x}\PY{o}{\PYZgt{}}\PY{p}{,} \PY{n}{cyclic}\PY{o}{\PYZlt{}}\PY{n}{y}\PY{o}{\PYZgt{}}\PY{o}{\PYZgt{}} 
      \PY{n}{slv}\PY{p}{(}\PY{n}{n}\PY{p}{[}\PY{n}{x}\PY{p}{]}\PY{p}{,} \PY{n}{n}\PY{p}{[}\PY{n}{y}\PY{p}{]}\PY{p}{)}\PY{p}{;} 
    \PY{n}{setup}\PY{p}{(}\PY{n}{slv}\PY{p}{,} \PY{n}{n}\PY{p}{)}\PY{p}{;} 
    \PY{n}{slv}\PY{p}{.}\PY{n}{solve}\PY{p}{(}\PY{n}{nt}\PY{p}{)}\PY{p}{;}
    \PY{n}{gp} \PY{o}{\PYZlt{}}\PY{o}{\PYZlt{}} \PY{l+s}{"}\PY{l+s}{set title 'mpdata\PYZlt{}}\PY{l+s}{"} \PY{o}{\PYZlt{}}\PY{o}{\PYZlt{}} \PY{n}{it} \PY{o}{\PYZlt{}}\PY{o}{\PYZlt{}} \PY{l+s}{"}\PY{l+s}{\PYZgt{} }\PY{l+s}{"}
       \PY{o}{\PYZlt{}}\PY{o}{\PYZlt{}} \PY{l+s}{"}\PY{l+s}{t=}\PY{l+s}{"} \PY{o}{\PYZlt{}}\PY{o}{\PYZlt{}} \PY{n}{nt} \PY{o}{\PYZlt{}}\PY{o}{\PYZlt{}} \PY{l+s}{"}\PY{l+s}{'}\PY{l+s+se}{\PYZbs{}n}\PY{l+s}{"}
       \PY{o}{\PYZlt{}}\PY{o}{\PYZlt{}} \PY{l+s}{"}\PY{l+s}{splot '-' binary}\PY{l+s}{"} \PY{o}{\PYZlt{}}\PY{o}{\PYZlt{}} \PY{n}{binfmt}
       \PY{o}{\PYZlt{}}\PY{o}{\PYZlt{}} \PY{l+s}{"}\PY{l+s}{with lines notitle}\PY{l+s+se}{\PYZbs{}n}\PY{l+s}{"}\PY{p}{;}
    \PY{n}{gp}\PY{p}{.}\PY{n}{sendBinary}\PY{p}{(}\PY{n}{slv}\PY{p}{.}\PY{n}{state}\PY{p}{(}\PY{p}{)}\PY{p}{.}\PY{n}{copy}\PY{p}{(}\PY{p}{)}\PY{p}{)}\PY{p}{;}
  \PY{p}{\PYZcb{}} 
  \PY{p}{\PYZob{}}
    \PY{k}{const} \PY{k+kt}{int} \PY{n}{it} \PY{o}{=} \PY{l+m+mi}{44}\PY{p}{;}
    \PY{n}{solver\PYZus{}mpdata}\PY{o}{\PYZlt{}}\PY{n}{it}\PY{p}{,} \PY{n}{cyclic}\PY{o}{\PYZlt{}}\PY{n}{x}\PY{o}{\PYZgt{}}\PY{p}{,} \PY{n}{cyclic}\PY{o}{\PYZlt{}}\PY{n}{y}\PY{o}{\PYZgt{}}\PY{o}{\PYZgt{}} 
      \PY{n}{slv}\PY{p}{(}\PY{n}{n}\PY{p}{[}\PY{n}{x}\PY{p}{]}\PY{p}{,} \PY{n}{n}\PY{p}{[}\PY{n}{y}\PY{p}{]}\PY{p}{)}\PY{p}{;} 
    \PY{n}{setup}\PY{p}{(}\PY{n}{slv}\PY{p}{,} \PY{n}{n}\PY{p}{)}\PY{p}{;} 
    \PY{n}{slv}\PY{p}{.}\PY{n}{solve}\PY{p}{(}\PY{n}{nt}\PY{p}{)}\PY{p}{;} 
    \PY{n}{gp} \PY{o}{\PYZlt{}}\PY{o}{\PYZlt{}} \PY{l+s}{"}\PY{l+s}{set title 'mpdata\PYZlt{}}\PY{l+s}{"} \PY{o}{\PYZlt{}}\PY{o}{\PYZlt{}} \PY{n}{it} \PY{o}{\PYZlt{}}\PY{o}{\PYZlt{}} \PY{l+s}{"}\PY{l+s}{\PYZgt{} }\PY{l+s}{"}
       \PY{o}{\PYZlt{}}\PY{o}{\PYZlt{}} \PY{l+s}{"}\PY{l+s}{t=}\PY{l+s}{"} \PY{o}{\PYZlt{}}\PY{o}{\PYZlt{}} \PY{n}{nt} \PY{o}{\PYZlt{}}\PY{o}{\PYZlt{}} \PY{l+s}{"}\PY{l+s}{'}\PY{l+s+se}{\PYZbs{}n}\PY{l+s}{"}
       \PY{o}{\PYZlt{}}\PY{o}{\PYZlt{}} \PY{l+s}{"}\PY{l+s}{splot '-' binary}\PY{l+s}{"} \PY{o}{\PYZlt{}}\PY{o}{\PYZlt{}} \PY{n}{binfmt}
       \PY{o}{\PYZlt{}}\PY{o}{\PYZlt{}} \PY{l+s}{"}\PY{l+s}{with lines notitle}\PY{l+s+se}{\PYZbs{}n}\PY{l+s}{"}\PY{p}{;}
    \PY{n}{gp}\PY{p}{.}\PY{n}{sendBinary}\PY{p}{(}\PY{n}{slv}\PY{p}{.}\PY{n}{state}\PY{p}{(}\PY{p}{)}\PY{p}{.}\PY{n}{copy}\PY{p}{(}\PY{p}{)}\PY{p}{)}\PY{p}{;}
  \PY{p}{\PYZcb{}}
\PY{p}{\PYZcb{}}
\PY{c+c1}{//listing21}
\end{Verbatim}
  \setcounter{linenocpp}{\value{FancyVerbLine}}%

  \section{Performance evaluation}\label{sec:perf}

  The three introduced implementations of MPDATA were tested with the following set-ups 
    employing free and open-source tools:
  \begin{description}
    \item[C++:]{~
      \begin{itemize}
        \item{GCC g++ 4.8.0\footnote{\label{fnt:gcc-snapshot}GNU Compiler Collection packaged in the Debian's gcc-snapshot\_20130222-1} 
          and Blitz++ 0.10}
        \item{LLVM Clang 3.2 and Blitz 0.10}
      \end{itemize}
    }
    \item[Python:]{~
      \begin{itemize}
        \item{CPython 2.7.3 and NumPy 1.7}
        \item{PyPy 1.9.0 with built-in NumPy implementation}
      \end{itemize}
    }
    \item[Fortran:]{~
      \begin{itemize}
        \item{GCC gfortran 4.8.0\textsuperscript{\ref{fnt:gcc-snapshot}}}
      \end{itemize}
    }
  \end{description}
  The performance tests were run on a Debian and an Ubuntu GNU/Linux systems with the above-listed software obtained 
    via binary packages from the distributions' package repositories (most recent package versions at the time of writing).
  The tests were performed on two 64-bit machines equipped with
    an AMD Phenom\textsuperscript{\texttrademark} II X6 1055T (800 MHz)
    and an Intel\textsuperscript{\textregistered} Core\textsuperscript{\texttrademark}~i5-2467M (1.6 GHz)
    processors.

  For both C++ and Fortran the GCC compilers were invoked with the \prog{-Ofast} and the 
    \prog{-march=native} options.
  The Clang compiler was invoked with the \prog{-O3}, the \prog{-mllvm -vectorize}, the \prog{-ffast-math} 
    and the \prog{-march=native} options.
  The CPython interpreter was invoked with the \prog{-OO} option.

  In addition to the standard Python implementation CPython,
    the Python code was tested with PyPy.
  PyPy is an alternative implementation of Python featuring a just-in-time compiler. 
  PyPy includes an experimental partial reimplementation of NumPy that compiles NumPy expressions into native assembler.
  Thanks to employment of lazy evaluation of array expressions (cf. Sect. \ref{sec:array})
    PyPy allows to eliminate the use of temporary matrices for storing intermediate results,
    and to perform multiple operations on the arrays within a single array index traversal
    \footnote{Lazy evaluation available in PyPy 1.9 has been temporarily removed from PyPy during a refactoring of
      the code. It'll be reinstantiated in the codebase as soon as possible, but past PyPy 2.0 release}.
  Consequently, PyPy allows to overcome the same performance-limiting factors as those addressed by Blitz++, although 
    the underlying mechanisms are different.
  In contrast to other solutions for improving performance of NumPy-based codes such as
    Cython\footnote{see \url{http://cython.org}}, 
    numexpr\footnote{see \url{http://code.google.com/p/numexpr/}} or 
    Numba\footnote{see \url{http://numba.pydata.org/}}, 
    PyPy does not require any modifications to the code.
  Thus, PyPy may serve as a drop-in replacement for CPython ready to be used with 
    previously-developed codes.
  
\begin{figure}[t]
    \center
    \includegraphics[height=.45\textwidth,angle=-90]{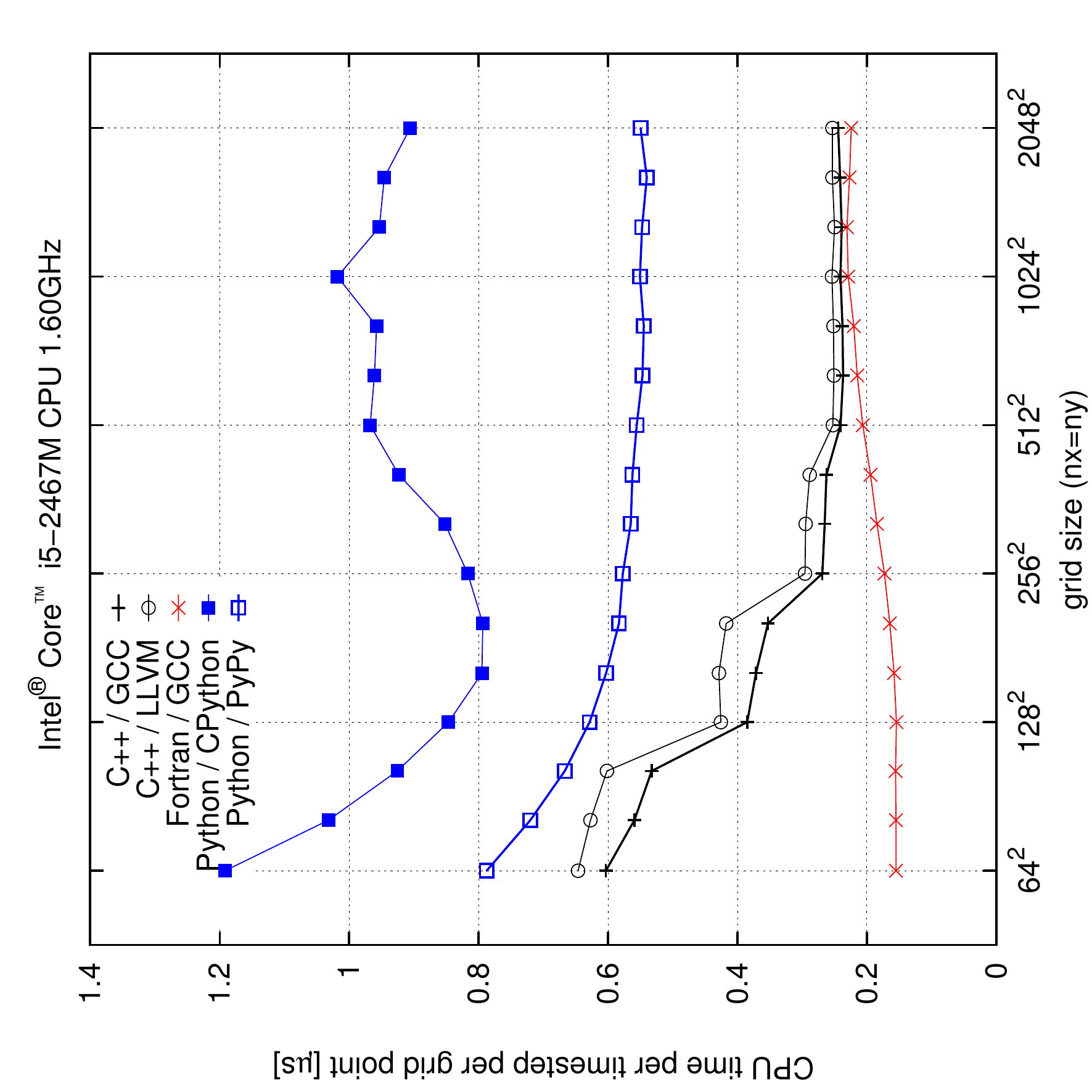}
    \caption{\label{fig:cpu-eyrie}
      Execution time statistics for the test runs described in Section~\ref{sec:perf}
        plotted as a function of grid size.
      Values of the total user mode CPU time reported by the GNU time utility are
        normalised by the grid size ($nx \cdot ny$) and the number of timesteps $nt=2^{24}/(nx \cdot ny)$.
      Before normalisation the time reported for an $nt=0$ run for a corresponding
        domain size is subtracted from the values.
      Both the $nt=0$ and $nt=2^{24}/()nx \cdot ny$ runs are repeated three times and
        only the shortest time is taken into account.
      Results obtained with an Intel\textsuperscript{\textregistered} 
        Core\textsuperscript{\texttrademark} i5 1.6 GHz processor.
    }
  \end{figure}

  The same set of tests was run with all four set-ups.
  Each test set consisted of 16 program runs.
  The test programs are analogous to the example code presented in section~\ref{sec:example}.
  The tests were run with different grid sizes ranging from 64$\times$64 to 2048$\times$2048.
  The Gaussian impulse was advected for $nt=2^{24}/(nx\cdot ny)$ timesteps ($2^{24}$ chosen arbitrarily), 
    in order to assure comparable timing accuracy for all grid sizes.
  Three MPDATA iterations were used (i.e. two corrective steps).
  The initial condition was loaded from a text file, and the final values were compared at the end of the test
    with values loaded from another text file assuring the same results were obtained with all four set-ups.
  The tests were run multiple times; program start-up, data loading, and output verification times were
    subtracted from the reported values (see caption of Figure~\ref{fig:cpu-eyrie} for details).

  \begin{figure}[t]
    \center
    \includegraphics[height=.45\textwidth,angle=-90]{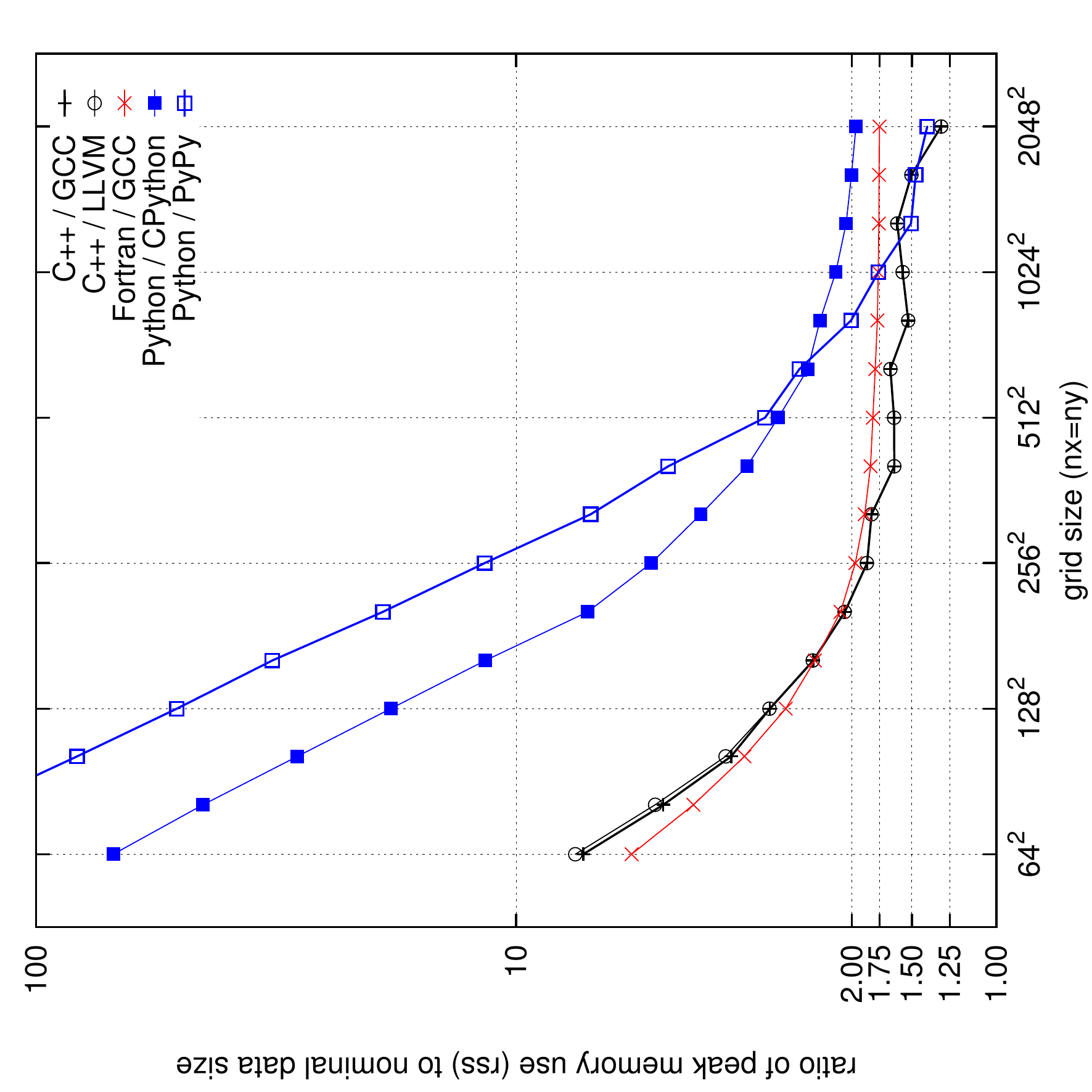}
    \caption{\label{fig:mem}
      Memory consumption statistics for the test runs described in Section~\ref{sec:perf}
        plotted as a function of grid size.
      Peak resident set size (rss) values reported by the GNU time utility are normalised by the size of
        data that needs to be allocated in the program to store all declared grid-sized arrays.
      Asymptotic values reached at the largest grid sizes are indicative 
        of temporary storage requirements.
    }
  \end{figure}

  Figure \ref{fig:mem} presents a plot of the peak memory use\footnote{The resident set size (rss)
    as reported by GNU time (version 1.7-24)} (identical for both considered CPUs)
    as a function of grid size.
  The plotted values are normalised by the nominal size of all data arrays used in the program
    (i.e. two (nx+2)$\times$(ny+2) arrays representing the two time levels of $\psi$, 
     a (nx+1)$\times$(ny+2) array representing the $C^{[x]}$ component of the Courant number field,
     a (nx+2)$\times$(ny+1) array representing the $C^{[y]}$ component, 
     and two pairs of arrays of the size of $C^{[x]}$ and $C^{[y]}$ for storing the 
     antidiffusive velocities, all composed of 8-byte double-precision floating point numbers).
  Plotted statistics reveal a notable memory footprint of the Python interpreter itself
    for both CPython and PyPy, losing its significance for domains larger than 1024$\times$1024.
  The roughly asymptotic values reached in all four set-ups for grid sizes larger that 1024$\times$1024
    are indicative of the amount of temporary memory used for array manipulation.
  PyPy- and Blitz++-based set-ups consume notably less memory than Fortran and CPython.
  This confirms the effectiveness of the just-in-time compilation (PyPy) and the expression-templates (Blitz++) techniques
    for elimination of temporary storage during array operations.

  \begin{figure}[t]
    \center
    \includegraphics[height=.45\textwidth,angle=-90]{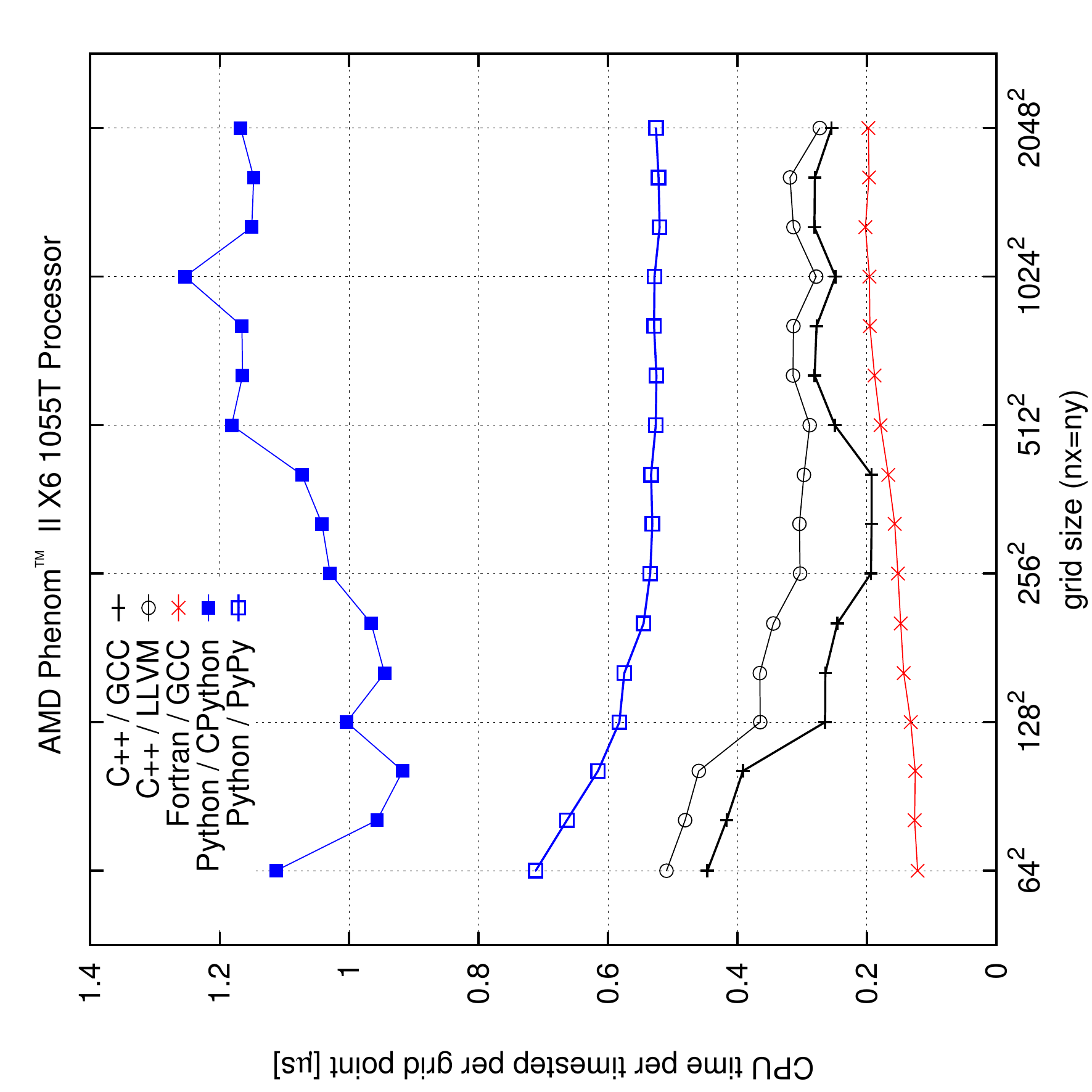}
    \caption{\label{fig:cpu-skua}
      Same as Fig.~\ref{fig:cpu-eyrie} for an AMD Phenom\textsuperscript{\texttrademark} II 800 MHz processor.
    }
  \end{figure}

  The CPU time statistics presented in Figures \ref{fig:cpu-eyrie} and \ref{fig:cpu-skua} reveal
    minor differences between results obtained with the two different processors.
  Presented results lead to the following observations
    (where by referring to language names, only the results obtained with the herein considered
     program codes, and software/hardware configurations are meant):
  \begin{itemize}
    \item{Fortran gives shortest execution times for any domain size;}
    \item{C++ execution times are less than twice those of Fortran for grids larger than 
      256$\times$256;}
    \item{CPython requires from around 4 to almost 10 times more CPU time than Fortran depending on the grid size;}
    \item{PyPy execution times are in most cases closer to C++ than to CPython.}
  \end{itemize}
  The support for OOP features in gfortran, the NumPy support in PyPy, and the relevant optimisation
    mechanisms in GCC are still in active development and hence the performance with some of the set-ups may 
    likely change with newer versions of these packages. 

  It is worth mentioning, that even though the three implementations are equally structured,
    the three considered languages have some inherent differences influencing the execution times.
  Notably, while Fortran and Blitz++ offer runtime array-bounds and array-shape checks as options
    not intended for use in production binaries, NumPy performs them always.
  Additionally, the C++ and Fortran set-ups may, in principle, benefit from GCC's auto-vectorisation
    features which do not have yet counterparts in CPython or PyPy.
  Finally, Fortran uses different ordering for storing array elements in memory, but since
    all tests were carried out using square grids, this should not have had any impact on the
    performance\footnote{Both Blitz++ and NumPy support Fortran's column-major ordering as well, 
    however this feature is still missing from PyPy's built-in NumPy implementation as of PyPy 1.9}.

  The authors do expect some performance gain could 
    be obtained by introducing into the codes some ''manual'' optimisations -- 
    code rearrangements aimed solely at the purpose of increasing performance.
  These were avoided intentionally as they degrade code readability,
    should in principle be handled by the compilers,
    and are generally advised to be avoided \citep[e.g.][section 3.12]{bib_CERNcpp}.

  \section{Discussion on the tradeoffs of language choice}

  One of the aims of this paper is to show the applicability of OOP features of the three
    programming languages (or language-library pairs) for scientific computing.
  The main focus is to represent
    what can be referred to as {\em blackboard abstractions} \citep{Rouson_et_al_2012} within the code.
  Presented benchmark tests, although quite simplistic, together with the experience gained 
    from the development of codes in three different
    languages provide a basis for discussion on the tradeoffs of programming language choice.
  The discussion concerns in principle the development of finite-difference solvers for 
    partial differential equations, but is likely applicable to the scientific software in general.
  A partly objective and partly subjective summary of pros and cons of C++, Python and Fortran
    is presented in the four following subsections.

  \subsection{OOP for blackboard abstractions}

  It was shown in section~\ref{sec:impl} that C++11/Blitz++, Python/NumPy and Fortran 2008
    provide comparable functionalities in terms of matching the blackboard abstractions
    within the program code.
  Taking into account solely the part of code representing particular formul\ae~
    (e.g. listings C.21, P.17, F.20 and equation \ref{eq:antidiff1}) all three
    languages allow to match (or surpass) \LaTeX~in its brevity of formula translation syntax.
  All three languages were shown to be capable of providing mechanisms to compactly represent such abstractions as:
  \begin{itemize}
    \item{loop-free array arithmetics;}
    \item{definitions of functions returning array-valued expressions;}
    \item{permutations of array indices allowing dimension-independent definitions
      of functions (see e.g. listings C.12 and C.13, P.10 and P.11, F.11 and F.12);}
    \item{fractional indexing of arrays corresponding to employment of a staggered grid.}
  \end{itemize}
  Three issues specific to Fortran that 
    resulted in employment of a more repetitive or cumbersome syntax than in C++ or Python
    were observed:
  \begin{itemize}
    \item{Fortran does not feature a mechanism allowing to reuse a single piece of code (algorithm)
      with different data types (compare e.g. listings C.6, P.5 and F.4) such as
      templates in C++ and the so-called duck typing in Python;}
    \item{Fortran does not allow function calls to appear on the left hand side
      of assignment (see e.g. how the \prog{ptr} pointers were used as a workaround in the \prog{cyclic\_fill\_halos}
      method in listing F.8);}
    \item{Fortran lacks support for arrays of arrays (cf. Sect. \ref{sec:sequence}).}
  \end{itemize}
  Interestingly, the limitation in extendability via inheritance was found to
    exist partially in NumPy as well (see footnote \ref{fnt:slice}).
  The lack of a counterpart in Fortran to the C++ template mechanism was identified in
    \citep{Cary_et_al_1997}
    as one of the key deficiencies of Fortran when compared with C++ in context 
    of applicability to object-oriented scientific programming.

  \subsection{Performance}
  
  The timing and memory usage statistics presented in figures \ref{fig:mem}-\ref{fig:cpu-skua}
    reveal that no single language/library/compiler set-up 
    corresponded to both shortest execution time and smallest memory footprint.

  One may consider performance measures addressing not only the program efficiency but also 
    the factors influencing the development and maintenance time/cost 
    \citep[of particular importance in scientific computing,][]{Wilson_2006}.
  Taking into account such measures as code length or coding time,
    the Python environment gains significantly.
  Presented Python code is shorter than the C++ and Fortran counterparts,
    and is simpler in terms of syntax and usage (see discussion below).

  Employment of the PyPy drop-in replacement for the standard Python implementation brings 
    Python's performance significantly closer to those of C++ and Fortran, in some
    cases making it the least memory consuming set-up.
  Python has already been the language of choice for scientific software projects having code clarity 
    or ease of use as the first requirement \citep[see e.g.][]{Barnes_and_Jones_2011}.
  PyPy's capability to improve performance of unmodified Python code may 
    make Python a favourable choice even if high performance is important, especially
    if a combined measure of performance and development cost is to be considered. 

  \subsection{Ease of use and abuse}

  Using the number of lines of code or the number of distinct language keywords
    needed to implement the MPDATA-based
    solver presented in section \ref{sec:impl} as measures of syntax 
    brevity, Python clearly surpasses its rivals.
  Python was developed with emphasis on code readability and object-orientation.
  Arguably, taking it to the extreme - Python uses line indentation to define 
    blocks of code and treats even single integers as objects.
  As a consequence Python is easy to learn and easy to teach.
  It is also much harder to abuse Python than C++ or Fortran
    (for instance with \prog{goto} statements, employment of the preprocessor,
    or the implicit typing in Fortran).

  Python implementations do not expose to the user the compilation or linking processes. 
  As a result, Python-written software is easier to deploy and share, especially 
    if multiple architectures and operating systems are targeted.
  However, there exist tools such as CMake\footnote{CMake is a family of open-source, cross-platform
    tools automating building, testing and packaging of C/C++/Fortran software,
    see \url{http://cmake.org/}}
    that allow to efficiently automate 
    building, testing and packaging of C++ and Fortran programs.

  Python is definitely easiest to debug among the three languages.
  Great debugging tools for C++ do exist, however the debugging and development is 
    often hindered by indecipherable compiler messages
    flooded with lengthy type names stemming from employment of templates.
  Support for the OOP features of Fortran among free and open source compilers, 
    debuggers and other programming aids remains immature.
    
  With both Fortran and Python, the memory footprint caused by employment
    of temporary objects in array arithmetics is dependant on compiler choice or
    the level of optimisations.
  In contrast, Blitz++ ensures temporary-array-free computations by design
    \citep{Veldhuizen_et_al_1997} avoiding unintentional performance loss.

  \subsection{Added values}

  The size of the programmers' community of a given language 
    influences the availability of trained personnel, 
    reusable software components and information resources.
  It also affects the maturity and quality of compilers and tools. 
  Fortran is a domain-specific language while Python and C++ are general-purpose languages
    with disproportionately larger users' communities.
  The OOP features of Fortran have not gained
    wide popularity among users \citep{Worth_2008}\footnote{An anecdotal yet significant
    example being the incomplete support for syntax-highlighting of modern Fortran in Vim and Emacs editors}.
  Fortran is no longer routinely taught at the universities \citep{Kendall_et_al_2008},
    in contrast to C++ and Python.
  An example of decreasing popularity of Fortran in academia 
    is the discontinuation of Fortran printed editions of the ''Numerical Recipes'' 
    series of Press et al.
 
  Blitz++ is one of several packages that offer high-performance object-oriented
    array manipulation functionality with C++ (and is not necessarily optimal for every
    purpose \citep{Iglberger_et_al_2012}).
  In contrast, the NumPy package became a de facto standard solution for Python.
  Consequently, numerous Python libraries adopted NumPy but
    there are apparently very few C++ libraries offering Blitz++ support out of the box
    (the gnuplot-iostream used in listing C.24 being a much-appreciated counterexample).
  However, Blitz++ allows to interface with virtually any library (including Fortran libraries), 
    by resorting to referencing the underlying memory with raw pointers.
 
  The availability and quality of libraries 
    that offer object-oriented interfaces
    differs among the three considered languages.
  The built-in standard libraries of Python and C++ are richer than
    those of Fortran and offer versatile data types, collections of
    algorithms and facilities for interaction with host operating system.
  In the authors' experience, the small popularity of OOP techniques among
    Fortran users is reflected in the library designs (including the Fortran's
    built-in library routines).
  What makes correct use of external libraries more difficult with Fortran
    is the lack of standard exception handling mechanism, a feature
    long and {\em much requested by the numerical community} \citep[][Foreword]{Press_et_al_1996}.

  Finally, the three languages differ as well with regard to availability of 
    mechanisms for leveraging shared-memory parallelisation (e.g. with multi-core processors).
  GCC supports OpenMP with Fortran and C++.
  The CPython and PyPy implementations of Python do not offer any
    built-in solution for multi-threading. 
  
  \section{Summary and outlook}

  Three implementations of a prototype solver 
    for the advection equation were introduced.
  The solvers are based on MPDATA - an algorithm of particular applicability
    in geophysical fluid dynamics \citep{Smolarkiewicz_2006}.
  All implementations follow the same object-oriented structure but are implemented
    in three different languages:
  \begin{itemize}
    \item{C++ with Blitz++;}
    \item{Python with NumPy;}
    \item{Fortran.}
  \end{itemize}

  Presented programs were developed making use of such recent
    developments as support for C++11 and Fortran 2008 in GCC, and
    the NumPy support in the PyPy implementation of Python.
  The fact that all considered standards are open and the employed
    tools implementing them are free and open-source
    is certainly an advantage \citep{Anel_2011}.

  The key conclusion is that all considered language/library/compiler
    set-ups offer possibilities for using OOP to compactly 
    represent the mathematical abstractions within the program code. 
  This creates the potential to improve code readability and brevity,
  \begin{itemize}
    \item{contributing to its 
      auditability, indispensable for credible and reproducible research in computational science 
      \citep{Post_et_al_2005, Merali_et_al_2010, Stodden_et_al_2012}; and
    }
    \item{helping to keep the programs maintainable and avoiding accumulation of the code 
      debt\footnote{See \citet{Buschmann_2011} for discussion of technical/code debt.} 
      that besets scientific software in such domains as climate modelling
      \citep{Freeman_et_al_2010}.
    }
  \end{itemize}
  \noindent 
  The performance evaluation revealed that:
    \begin{itemize}
      \item{the Fortran set-up offered shortest execution times,}
      \item{it took the C++ set-up less than twice longer to compute than Fortran,}
      \item{C++ and PyPy set-ups offered significantly smaller memory consumption 
        than Fortran and CPython for larger domains,}
      \item{the PyPy set-up was roughly twice slower than C++ and up to twice faster than CPython.}
    \end{itemize}
  The three equally-structured implementations required ca. 200, 300, and 500 lines of code 
    in Python, C++ and Fortran, respectively.  

  In addition to the source code presented within the text,
    a set of tests and build-/test-automation scripts
    allowing to reproduce the analysis and plots presented in section~\ref{sec:perf} are all 
    available in the CPC Program Library and at
    the project repository\footnote{git repository at \url{http://github.com/slayoo/mpdata/}},
    and are released under the GNU GPL license \citep{GPLv3}.
  The authors encourage to use the presented codes for teaching and benchmarking purposes.

  The OOP design enhances the possibilities to reuse and extend the presented code.
  Development is underway of an object-oriented C++ library featuring concepts presented herein,
    supporting integration in one to three dimensions, handling systems of equations with source terms, 
    providing miscellaneous options of MPDATA and several parallel processing approaches.

  \section*{Acknowledgements}
  {\small
    \noindent 
    We thank Piotr Smolarkiewicz and Hanna Pawłowska for their help throughout the project.
    This study was partly inspired by the lectures of Lech Łobocki.

    \noindent
    Tobias Burnus, Julian Cummings, Ond\v{r}ej \v{C}ert\'ik, Patrik Jonsson,
      Arjen Markus, Zbigniew Piotrowski, Davide del Vento and Janus Weil 
      provided valuable feedback to the initial version of the
      manuscript and/or responses to questions posted to Blitz++ and gfortran mailing lists.
    
    \noindent
    SA, AJ and DJ acknowledge funding from the Polish National Science Centre
      (project no. 2011/01/N/ST10/01483).

    \noindent
    Part of the work was carried out during a visit of SA to the National
      Center for Atmospheric Research (NCAR) in Boulder, Colorado, USA.
    NCAR is operated by the University Corporation for Atmospheric Research.
    The visit was funded by the Foundation for Polish Science (START programme).
 
    \noindent
    Development of NumPy support in PyPy was led by Alex Gaynor, Matti Picus and MF.
  }

  \clearpage
  \appendix

  \setcounter{section}{15} 
  \section{Python code for sections \ref{sec:cyclic}--\ref{sec:mpdatasolver}}\label{app:P}

  \subsection*{{\bf Periodic Boundaries} (cf. Sect. \ref{sec:cyclic})}
  \noindent 
  \addtocounter{lstnopyt}{1}%
  \renewcommand*\FancyVerbStartString{\PY{c}{\PYZsh{}listing08}}
  \renewcommand*\FancyVerbStopString{\PY{c}{\PYZsh{}listing09}}
  \setcounter{FancyVerbLine}{\thelinenopyt}%
  \fvset{label={listing~P.\thelstnopyt~(Python)},rulecolor=\color{blue},stepnumber=1}%
  \setcounter{linenopyt}{\value{FancyVerbLine}}%

  \subsection*{{\bf Donor-cell formul\ae}~(cf. Sect. \ref{sec:donor})}
  \addtocounter{lstnopyt}{1}%
  \renewcommand*\FancyVerbStartString{\PY{c}{\PYZsh{}listing09}}
  \renewcommand*\FancyVerbStopString{\PY{c}{\PYZsh{}listing10}}
  \setcounter{FancyVerbLine}{\thelinenopyt}%
  \fvset{label={listing~P.\thelstnopyt~(Python)},rulecolor=\color{blue},stepnumber=1}%
  \setcounter{linenopyt}{\value{FancyVerbLine}}%

  \addtocounter{lstnopyt}{1}%
  \renewcommand*\FancyVerbStartString{\PY{c}{\PYZsh{}listing10}}
  \renewcommand*\FancyVerbStopString{\PY{c}{\PYZsh{}listing11}}
  \setcounter{FancyVerbLine}{\thelinenopyt}%
  \fvset{label={listing~P.\thelstnopyt~(Python)},rulecolor=\color{blue},stepnumber=1}%
  \setcounter{linenopyt}{\value{FancyVerbLine}}%

  \addtocounter{lstnopyt}{1}%
  \renewcommand*\FancyVerbStartString{\PY{c}{\PYZsh{}listing11}}
  \renewcommand*\FancyVerbStopString{\PY{c}{\PYZsh{}listing12}}
  \setcounter{FancyVerbLine}{\thelinenopyt}%
  \fvset{label={listing~P.\thelstnopyt~(Python)},rulecolor=\color{blue},stepnumber=1}%
  \setcounter{linenopyt}{\value{FancyVerbLine}}%

  \subsection*{{\bf Donor-cell solver} (cf. Sect. \ref{sec:donorcell_solver})}
  \addtocounter{lstnopyt}{1}%
  \renewcommand*\FancyVerbStartString{\PY{c}{\PYZsh{}listing12}}
  \renewcommand*\FancyVerbStopString{\PY{c}{\PYZsh{}listing13}}
  \setcounter{FancyVerbLine}{\thelinenopyt}%
  \fvset{label={listing~P.\thelstnopyt~(Python)},rulecolor=\color{blue},stepnumber=1}%
  \setcounter{linenopyt}{\value{FancyVerbLine}}%

  \subsection*{{\bf MPDATA formul\ae}~(cf. Sect. \ref{sec:mpdata})}
  \addtocounter{lstnopyt}{1}%
  \renewcommand*\FancyVerbStartString{\PY{c}{\PYZsh{}listing13}}
  \renewcommand*\FancyVerbStopString{\PY{c}{\PYZsh{}listing14}}
  \setcounter{FancyVerbLine}{\thelinenopyt}%
  \fvset{label={listing~P.\thelstnopyt~(Python)},rulecolor=\color{blue},stepnumber=1}%
  \setcounter{linenopyt}{\value{FancyVerbLine}}%

  \addtocounter{lstnopyt}{1}%
  \renewcommand*\FancyVerbStartString{\PY{c}{\PYZsh{}listing14}}
  \renewcommand*\FancyVerbStopString{\PY{c}{\PYZsh{}listing15}}
  \setcounter{FancyVerbLine}{\thelinenopyt}%
  \fvset{label={listing~P.\thelstnopyt~(Python)},rulecolor=\color{blue},stepnumber=1}%
  \setcounter{linenopyt}{\value{FancyVerbLine}}%

  \addtocounter{lstnopyt}{1}%
  \renewcommand*\FancyVerbStartString{\PY{c}{\PYZsh{}listing15}}
  \renewcommand*\FancyVerbStopString{\PY{c}{\PYZsh{}listing16}}
  \setcounter{FancyVerbLine}{\thelinenopyt}%
  \fvset{label={listing~P.\thelstnopyt~(Python)},rulecolor=\color{blue},stepnumber=1}%
  \setcounter{linenopyt}{\value{FancyVerbLine}}%

  \addtocounter{lstnopyt}{1}%
  \renewcommand*\FancyVerbStartString{\PY{c}{\PYZsh{}listing16}}
  \renewcommand*\FancyVerbStopString{\PY{c}{\PYZsh{}listing17}}
  \setcounter{FancyVerbLine}{\thelinenopyt}%
  \fvset{label={listing~P.\thelstnopyt~(Python)},rulecolor=\color{blue},stepnumber=1}%
  \setcounter{linenopyt}{\value{FancyVerbLine}}%

  \addtocounter{lstnopyt}{1}%
  \renewcommand*\FancyVerbStartString{\PY{c}{\PYZsh{}listing17}}
  \renewcommand*\FancyVerbStopString{\PY{c}{\PYZsh{}listing18}}
  \setcounter{FancyVerbLine}{\thelinenopyt}%
  \fvset{label={listing~P.\thelstnopyt~(Python)},rulecolor=\color{blue},stepnumber=1}%
  \setcounter{linenopyt}{\value{FancyVerbLine}}%

  \subsection*{{\bf An MPDATA solver} (cf. Sect. \ref{sec:mpdatasolver})}
  \addtocounter{lstnopyt}{1}%
  \renewcommand*\FancyVerbStartString{\PY{c}{\PYZsh{}listing18}}
  \renewcommand*\FancyVerbStopString{\PY{c}{\PYZsh{}listing19}}
  \setcounter{FancyVerbLine}{\thelinenopyt}%
  \fvset{label={listing~P.\thelstnopyt~(Python)},rulecolor=\color{blue},stepnumber=1}%
  \setcounter{linenopyt}{\value{FancyVerbLine}}%

  \subsection*{{\bf Usage example} (cf. Sect. \ref{sec:example})}
  \addtocounter{lstnopyt}{1}%
  \renewcommand*\FancyVerbStartString{\PY{c}{\PYZsh{}listing19}}
  \renewcommand*\FancyVerbStopString{\PY{c}{\PYZsh{}listing20}}
  \setcounter{FancyVerbLine}{\thelinenopyt}%
  \fvset{label={listing~P.\thelstnopyt~(Python)},rulecolor=\color{blue},stepnumber=1}%
  \begin{Verbatim}[commandchars=\\\{\}]
\PY{k+kn}{from} \PY{n+nn}{listings} \PY{k+kn}{import} \PY{o}{*}
\PY{k+kn}{import} \PY{n+nn}{sys}

\PY{k}{def} \PY{n+nf}{read\PYZus{}file}\PY{p}{(}\PY{n}{fname}\PY{p}{,} \PY{n}{nx}\PY{p}{,} \PY{n}{ny}\PY{p}{)}\PY{p}{:}
  \PY{n}{tmp} \PY{o}{=} \PY{n}{numpy}\PY{o}{.}\PY{n}{empty}\PY{p}{(}\PY{p}{(}\PY{n}{nx}\PY{p}{,} \PY{n}{ny}\PY{p}{)}\PY{p}{,} \PY{n}{real\PYZus{}t}\PY{p}{)}
  \PY{k}{with} \PY{n+nb}{open}\PY{p}{(}\PY{n}{fname}\PY{p}{,} \PY{l+s}{'}\PY{l+s}{r}\PY{l+s}{'}\PY{p}{)} \PY{k}{as} \PY{n}{f}\PY{p}{:}
    \PY{n}{x} \PY{o}{=} \PY{l+m+mi}{0}
    \PY{k}{for} \PY{n}{line} \PY{o+ow}{in} \PY{n}{f}\PY{p}{:}
      \PY{n}{tmp}\PY{p}{[}\PY{n}{x}\PY{p}{,}\PY{p}{:}\PY{p}{]} \PY{o}{=} \PY{n}{numpy}\PY{o}{.}\PY{n}{fromstring}\PY{p}{(}\PY{n}{line}\PY{p}{,} \PY{n}{dtype}\PY{o}{=}\PY{n}{real\PYZus{}t}\PY{p}{,} \PY{n}{sep}\PY{o}{=}\PY{l+s}{'}\PY{l+s+se}{\PYZbs{}t}\PY{l+s}{'}\PY{p}{)}
      \PY{n}{x} \PY{o}{+}\PY{o}{=} \PY{l+m+mi}{1}
  \PY{k}{assert}\PY{p}{(}\PY{n}{x} \PY{o}{==} \PY{n}{nx}\PY{p}{)}
  \PY{k}{return} \PY{n}{tmp}

\PY{k}{def} \PY{n+nf}{main}\PY{p}{(}\PY{p}{)}\PY{p}{:}
  \PY{k}{if} \PY{p}{(}\PY{n+nb}{len}\PY{p}{(}\PY{n}{sys}\PY{o}{.}\PY{n}{argv}\PY{p}{)} \PY{o}{!=} \PY{p}{(}\PY{l+m+mi}{9} \PY{o}{+} \PY{l+m+mi}{1}\PY{p}{)}\PY{p}{)} \PY{p}{:} 
    \PY{k}{raise} \PY{n+ne}{Exception}\PY{p}{(}\PY{l+s}{'}\PY{l+s}{expecting 9 arguments (nx, ny, Cx, Cy, nt, it, f.in, f.out, dec)}\PY{l+s}{'}\PY{p}{)}

  \PY{n}{nx} \PY{o}{=} \PY{n+nb}{int}\PY{p}{(}\PY{n}{sys}\PY{o}{.}\PY{n}{argv}\PY{p}{[}\PY{l+m+mi}{1}\PY{p}{]}\PY{p}{)}
  \PY{n}{ny} \PY{o}{=} \PY{n+nb}{int}\PY{p}{(}\PY{n}{sys}\PY{o}{.}\PY{n}{argv}\PY{p}{[}\PY{l+m+mi}{2}\PY{p}{]}\PY{p}{)}
  \PY{n}{Cx} \PY{o}{=} \PY{n+nb}{float}\PY{p}{(}\PY{n}{sys}\PY{o}{.}\PY{n}{argv}\PY{p}{[}\PY{l+m+mi}{3}\PY{p}{]}\PY{p}{)}
  \PY{n}{Cy} \PY{o}{=} \PY{n+nb}{float}\PY{p}{(}\PY{n}{sys}\PY{o}{.}\PY{n}{argv}\PY{p}{[}\PY{l+m+mi}{4}\PY{p}{]}\PY{p}{)}
  \PY{n}{nt} \PY{o}{=} \PY{n+nb}{int}\PY{p}{(}\PY{n}{sys}\PY{o}{.}\PY{n}{argv}\PY{p}{[}\PY{l+m+mi}{5}\PY{p}{]}\PY{p}{)}
  \PY{n}{it} \PY{o}{=} \PY{n+nb}{int}\PY{p}{(}\PY{n}{sys}\PY{o}{.}\PY{n}{argv}\PY{p}{[}\PY{l+m+mi}{6}\PY{p}{]}\PY{p}{)}
  \PY{n}{fname} \PY{o}{=} \PY{n}{sys}\PY{o}{.}\PY{n}{argv}\PY{p}{[}\PY{l+m+mi}{7}\PY{p}{]}
  \PY{n}{fout} \PY{o}{=} \PY{n}{sys}\PY{o}{.}\PY{n}{argv}\PY{p}{[}\PY{l+m+mi}{8}\PY{p}{]}
  \PY{n}{dec} \PY{o}{=} \PY{n+nb}{int}\PY{p}{(}\PY{n}{sys}\PY{o}{.}\PY{n}{argv}\PY{p}{[}\PY{l+m+mi}{9}\PY{p}{]}\PY{p}{)}

\PY{c}{\PYZsh{}listing19}
  \PY{n}{slv} \PY{o}{=} \PY{n}{solver\PYZus{}mpdata}\PY{p}{(}\PY{n}{it}\PY{p}{,} \PY{n}{cyclic}\PY{p}{,} \PY{n}{cyclic}\PY{p}{,} \PY{n}{nx}\PY{p}{,} \PY{n}{ny}\PY{p}{)}
  \PY{n}{slv}\PY{o}{.}\PY{n}{state}\PY{p}{(}\PY{p}{)}\PY{p}{[}\PY{p}{:}\PY{p}{]} \PY{o}{=} \PY{n}{read\PYZus{}file}\PY{p}{(}\PY{n}{fname}\PY{p}{,} \PY{n}{nx}\PY{p}{,} \PY{n}{ny}\PY{p}{)}
  \PY{n}{slv}\PY{o}{.}\PY{n}{courant}\PY{p}{(}\PY{l+m+mi}{0}\PY{p}{)}\PY{p}{[}\PY{p}{:}\PY{p}{]} \PY{o}{=} \PY{n}{Cx}
  \PY{n}{slv}\PY{o}{.}\PY{n}{courant}\PY{p}{(}\PY{l+m+mi}{1}\PY{p}{)}\PY{p}{[}\PY{p}{:}\PY{p}{]} \PY{o}{=} \PY{n}{Cy}
  \PY{n}{slv}\PY{o}{.}\PY{n}{solve}\PY{p}{(}\PY{n}{nt}\PY{p}{)}
\PY{c}{\PYZsh{}listing20}

  \PY{n}{diff} \PY{o}{=} \PY{n}{numpy}\PY{o}{.}\PY{n}{amax}\PY{p}{(}\PY{n+nb}{abs}\PY{p}{(}\PY{n}{slv}\PY{o}{.}\PY{n}{state}\PY{p}{(}\PY{p}{)} \PY{o}{-} \PY{n}{read\PYZus{}file}\PY{p}{(}\PY{n}{fout}\PY{p}{,} \PY{n}{nx}\PY{p}{,} \PY{n}{ny}\PY{p}{)}\PY{p}{)}\PY{p}{)}
  \PY{k}{print} \PY{l+s}{"}\PY{l+s}{diff=}\PY{l+s}{"}\PY{p}{,} \PY{n}{diff}
  \PY{k}{if} \PY{p}{(}\PY{n}{diff} \PY{o}{\PYZgt{}}\PY{o}{=} \PY{o}{.}\PY{l+m+mi}{5} \PY{o}{*} \PY{n+nb}{pow}\PY{p}{(}\PY{l+m+mi}{10}\PY{p}{,} \PY{o}{-}\PY{n}{dec}\PY{p}{)}\PY{p}{)}\PY{p}{:} 
    \PY{k}{print} \PY{n}{slv}\PY{o}{.}\PY{n}{state}\PY{p}{(}\PY{p}{)}\PY{o}{.}\PY{n}{dtype}\PY{p}{,} \PY{n}{slv}\PY{o}{.}\PY{n}{state}\PY{p}{(}\PY{p}{)}
    \PY{n}{tmp} \PY{o}{=} \PY{n}{read\PYZus{}file}\PY{p}{(}\PY{n}{fout}\PY{p}{,} \PY{n}{nx}\PY{p}{,} \PY{n}{ny}\PY{p}{)}
    \PY{k}{print} \PY{n}{tmp}\PY{o}{.}\PY{n}{dtype}\PY{p}{,} \PY{n}{tmp}
    \PY{k}{print} \PY{n}{numpy}\PY{o}{.}\PY{n}{max}\PY{p}{(}\PY{n}{numpy}\PY{o}{.}\PY{n}{abs}\PY{p}{(}\PY{n}{slv}\PY{o}{.}\PY{n}{state}\PY{p}{(}\PY{p}{)} \PY{o}{-} \PY{n}{tmp}\PY{p}{)}\PY{p}{)}
    \PY{k}{raise} \PY{n+ne}{Exception}\PY{p}{(}\PY{p}{)}

\PY{k}{if} \PY{n}{\PYZus{}\PYZus{}name\PYZus{}\PYZus{}} \PY{o}{==} \PY{l+s}{"}\PY{l+s}{\PYZus{}\PYZus{}main\PYZus{}\PYZus{}}\PY{l+s}{"}\PY{p}{:}
    \PY{n}{main}\PY{p}{(}\PY{p}{)}
\end{Verbatim}
  \setcounter{linenopyt}{\value{FancyVerbLine}}%

  \setcounter{section}{5} 
  \section{Fortran code for sections \ref{sec:cyclic}--\ref{sec:mpdatasolver}}\label{app:F}

  \subsection*{{\bf Periodic boundaries} (cf. Sect. \ref{sec:cyclic})}
  \addtocounter{lstnofor}{1}%
  \renewcommand*\FancyVerbStartString{\PY{c}{!listing08}}
  \renewcommand*\FancyVerbStopString{\PY{c}{!listing09}}
  \setcounter{FancyVerbLine}{\thelinenofor}%
  \fvset{label={listing~F.\thelstnofor~(Fortran)},rulecolor=\color{red},stepnumber=1}%
  \setcounter{linenofor}{\value{FancyVerbLine}}%

  \subsection*{{\bf Donor-cell formul\ae}~(cf. Sect. \ref{sec:donor})}
  \addtocounter{lstnofor}{1}%
  \renewcommand*\FancyVerbStartString{\PY{c}{!listing09}}
  \renewcommand*\FancyVerbStopString{\PY{c}{!listing10}}
  \setcounter{FancyVerbLine}{\thelinenofor}%
  \fvset{label={listing~F.\thelstnofor~(Fortran)},rulecolor=\color{red},stepnumber=1}%
  \setcounter{linenofor}{\value{FancyVerbLine}}%

  \addtocounter{lstnofor}{1}%
  \renewcommand*\FancyVerbStartString{\PY{c}{!listing10}}
  \renewcommand*\FancyVerbStopString{\PY{c}{!listing11}}
  \setcounter{FancyVerbLine}{\thelinenofor}%
  \fvset{label={listing~F.\thelstnofor~(Fortran)},rulecolor=\color{red},stepnumber=1}%
  \setcounter{linenofor}{\value{FancyVerbLine}}%

  \addtocounter{lstnofor}{1}%
  \renewcommand*\FancyVerbStartString{\PY{c}{!listing11}}
  \renewcommand*\FancyVerbStopString{\PY{c}{!listing12}}
  \setcounter{FancyVerbLine}{\thelinenofor}%
  \fvset{label={listing~F.\thelstnofor~(Fortran)},rulecolor=\color{red},stepnumber=1}%
  \setcounter{linenofor}{\value{FancyVerbLine}}%

  \addtocounter{lstnofor}{1}%
  \renewcommand*\FancyVerbStartString{\PY{c}{!listing12}}
  \renewcommand*\FancyVerbStopString{\PY{c}{!listing13}}
  \setcounter{FancyVerbLine}{\thelinenofor}%
  \fvset{label={listing~F.\thelstnofor~(Fortran)},rulecolor=\color{red},stepnumber=1}%
  \setcounter{linenofor}{\value{FancyVerbLine}}%

  \addtocounter{lstnofor}{1}%
  \renewcommand*\FancyVerbStartString{\PY{c}{!listing13}}
  \renewcommand*\FancyVerbStopString{\PY{c}{!listing14}}
  \setcounter{FancyVerbLine}{\thelinenofor}%
  \fvset{label={listing~F.\thelstnofor~(Fortran)},rulecolor=\color{red},stepnumber=1}%
  \setcounter{linenofor}{\value{FancyVerbLine}}%

  \subsection*{{\bf Donor-cell solver} (cf. Sect. \ref{sec:donorcell_solver})}
  \addtocounter{lstnofor}{1}%
  \renewcommand*\FancyVerbStartString{\PY{c}{!listing14}}
  \renewcommand*\FancyVerbStopString{\PY{c}{!listing15}}
  \setcounter{FancyVerbLine}{\thelinenofor}%
  \fvset{label={listing~F.\thelstnofor~(Fortran)},rulecolor=\color{red},stepnumber=1}%
  \setcounter{linenofor}{\value{FancyVerbLine}}%

  \subsection*{{\bf MPDATA formul\ae}~(cf. Sect. \ref{sec:mpdata})}
  \addtocounter{lstnofor}{1}%
  \renewcommand*\FancyVerbStartString{\PY{c}{!listing15}}
  \renewcommand*\FancyVerbStopString{\PY{c}{!listing16}}
  \setcounter{FancyVerbLine}{\thelinenofor}%
  \fvset{label={listing~F.\thelstnofor~(Fortran)},rulecolor=\color{red},stepnumber=1}%
  \setcounter{linenofor}{\value{FancyVerbLine}}%

  \addtocounter{lstnofor}{1}%
  \renewcommand*\FancyVerbStartString{\PY{c}{!listing16}}
  \renewcommand*\FancyVerbStopString{\PY{c}{!listing17}}
  \setcounter{FancyVerbLine}{\thelinenofor}%
  \fvset{label={listing~F.\thelstnofor~(Fortran)},rulecolor=\color{red},stepnumber=1}%
  \setcounter{linenofor}{\value{FancyVerbLine}}%

  \addtocounter{lstnofor}{1}%
  \renewcommand*\FancyVerbStartString{\PY{c}{!listing17}}
  \renewcommand*\FancyVerbStopString{\PY{c}{!listing18}}
  \setcounter{FancyVerbLine}{\thelinenofor}%
  \fvset{label={listing~F.\thelstnofor~(Fortran)},rulecolor=\color{red},stepnumber=1}%
  \setcounter{linenofor}{\value{FancyVerbLine}}%

  \addtocounter{lstnofor}{1}%
  \renewcommand*\FancyVerbStartString{\PY{c}{!listing18}}
  \renewcommand*\FancyVerbStopString{\PY{c}{!listing19}}
  \setcounter{FancyVerbLine}{\thelinenofor}%
  \fvset{label={listing~F.\thelstnofor~(Fortran)},rulecolor=\color{red},stepnumber=1}%
  \setcounter{linenofor}{\value{FancyVerbLine}}%

  \addtocounter{lstnofor}{1}%
  \renewcommand*\FancyVerbStartString{\PY{c}{!listing19}}
  \renewcommand*\FancyVerbStopString{\PY{c}{!listing20}}
  \setcounter{FancyVerbLine}{\thelinenofor}%
  \fvset{label={listing~F.\thelstnofor~(Fortran)},rulecolor=\color{red},stepnumber=1}%
  \setcounter{linenofor}{\value{FancyVerbLine}}%

  \addtocounter{lstnofor}{1}%
  \renewcommand*\FancyVerbStartString{\PY{c}{!listing20}}
  \renewcommand*\FancyVerbStopString{\PY{c}{!listing21}}
  \setcounter{FancyVerbLine}{\thelinenofor}%
  \fvset{label={listing~F.\thelstnofor~(Fortran)},rulecolor=\color{red},stepnumber=1}%
  \setcounter{linenofor}{\value{FancyVerbLine}}%

  \addtocounter{lstnofor}{1}%
  \renewcommand*\FancyVerbStartString{\PY{c}{!listing21}}
  \renewcommand*\FancyVerbStopString{\PY{c}{!listing22}}
  \setcounter{FancyVerbLine}{\thelinenofor}%
  \fvset{label={listing~F.\thelstnofor~(Fortran)},rulecolor=\color{red},stepnumber=1}%
  \setcounter{linenofor}{\value{FancyVerbLine}}%

  \subsection*{{\bf An MPDATA solver} (cf. Sect. \ref{sec:mpdatasolver})}
  \addtocounter{lstnofor}{1}%
  \renewcommand*\FancyVerbStartString{\PY{c}{!listing22}}
  \renewcommand*\FancyVerbStopString{\PY{c}{!listing23}}
  \setcounter{FancyVerbLine}{\thelinenofor}%
  \fvset{label={listing~F.\thelstnofor~(Fortran)},rulecolor=\color{red},stepnumber=1}%
  \setcounter{linenofor}{\value{FancyVerbLine}}%

  \subsection*{{\bf Usage example} (cf. Sect. \ref{sec:example})}
  \addtocounter{lstnofor}{1}%
  \renewcommand*\FancyVerbStartString{\PY{c}{!listing23}}
  \renewcommand*\FancyVerbStopString{\PY{c}{!listing24}}
  \setcounter{FancyVerbLine}{\thelinenofor}%
  \fvset{label={listing~F.\thelstnofor~(Fortran)},rulecolor=\color{red},stepnumber=1}%
  \begin{Verbatim}[commandchars=\\\{\}]
\PY{k}{program }\PY{n+nv}{test}
  \PY{k}{use }\PY{n+nv}{solver\PYZus{}mpdata\PYZus{}m}
  \PY{k}{use }\PY{n+nv}{cyclic\PYZus{}m}
  \PY{k}{implicit }\PY{k}{none}

\PY{k}{  }\PY{k}{if} \PY{p}{(}\PY{n+nb}{command\PYZus{}argument\PYZus{}count}\PY{p}{(}\PY{p}{)} \PY{o}{/}\PY{o}{=} \PY{p}{(}\PY{l+m+mi}{9}\PY{p}{)}\PY{p}{)} \PY{k}{then}
\PY{k}{    }\PY{k}{print}\PY{o}{*}\PY{p}{,} \PY{l+s+s2}{"expecting 9 arguments (nx, ny, Cx, Cy, nt, it, f.in, f.out, dec)"}
    \PY{k}{stop }\PY{l+m+mi}{1}
  \PY{k}{end }\PY{k}{if}

\PY{k}{  }\PY{k}{block}
\PY{c}{!listing23}
    \PY{k}{type}\PY{p}{(}\PY{n+nv}{mpdata\PYZus{}t}\PY{p}{)} \PY{k+kd}{::} \PY{n+nv}{slv}
    \PY{k}{type}\PY{p}{(}\PY{n+nv}{cyclic\PYZus{}t}\PY{p}{)}\PY{p}{,} \PY{k}{target} \PY{k+kd}{::} \PY{n+nv}{bcx}\PY{p}{,} \PY{n+nv}{bcy}
    \PY{k+kt}{integer} \PY{k+kd}{::} \PY{n+nv}{nx}\PY{p}{,} \PY{n+nv}{ny}\PY{p}{,} \PY{n+nv}{nt}\PY{p}{,} \PY{n+nv}{it}
    \PY{k+kt}{real}\PY{p}{(}\PY{n+nv}{real\PYZus{}t}\PY{p}{)} \PY{k+kd}{::} \PY{n+nv}{Cx}\PY{p}{,} \PY{n+nv}{Cy}
    \PY{k+kt}{real}\PY{p}{(}\PY{n+nv}{real\PYZus{}t}\PY{p}{)}\PY{p}{,} \PY{k}{pointer} \PY{k+kd}{::} \PY{n+nv}{ptr}\PY{p}{(}\PY{p}{:}\PY{p}{,}\PY{p}{:}\PY{p}{)}
\PY{c}{!listing24}
    \PY{k+kt}{character} \PY{p}{(}\PY{n+nb}{len}\PY{o}{=}\PY{l+m+mi}{666}\PY{p}{)} \PY{k+kd}{::} \PY{n+nv}{fname}
    \PY{k+kt}{integer} \PY{k+kd}{::} \PY{n+nv}{dec}

    \PY{n+nv}{nx} \PY{o}{=} \PY{n+nv}{arg2int}\PY{p}{(}\PY{l+m+mi}{1}\PY{p}{)}
    \PY{n+nv}{ny} \PY{o}{=} \PY{n+nv}{arg2int}\PY{p}{(}\PY{l+m+mi}{2}\PY{p}{)}
    \PY{n+nv}{Cx} \PY{o}{=} \PY{n+nv}{arg2real}\PY{p}{(}\PY{l+m+mi}{3}\PY{p}{)}
    \PY{n+nv}{Cy} \PY{o}{=} \PY{n+nv}{arg2real}\PY{p}{(}\PY{l+m+mi}{4}\PY{p}{)}
    \PY{n+nv}{nt} \PY{o}{=} \PY{n+nv}{arg2int}\PY{p}{(}\PY{l+m+mi}{5}\PY{p}{)}
    \PY{n+nv}{it} \PY{o}{=} \PY{n+nv}{arg2int}\PY{p}{(}\PY{l+m+mi}{6}\PY{p}{)}
    \PY{n+nv}{dec} \PY{o}{=} \PY{n+nv}{arg2int}\PY{p}{(}\PY{l+m+mi}{9}\PY{p}{)}
    \PY{k}{call }\PY{n+nb}{get\PYZus{}command\PYZus{}argument}\PY{p}{(}\PY{l+m+mi}{7}\PY{p}{,} \PY{n+nv}{fname}\PY{p}{)}

\PY{c}{!listing25}
    \PY{k}{call }\PY{n+nv}{slv}\PY{p}{\PYZpc{}}\PY{n+nv}{ctor}\PY{p}{(}\PY{n+nv}{it}\PY{p}{,} \PY{n+nv}{bcx}\PY{p}{,} \PY{n+nv}{bcy}\PY{p}{,} \PY{n+nv}{nx}\PY{p}{,} \PY{n+nv}{ny}\PY{p}{)}

    \PY{n+nv}{ptr} \PY{o}{=}\PY{o}{\PYZgt{}} \PY{n+nv}{slv}\PY{p}{\PYZpc{}}\PY{n+nv}{state}\PY{p}{(}\PY{p}{)} 
    \PY{k}{call }\PY{n+nv}{read\PYZus{}file}\PY{p}{(}\PY{n+nv}{fname}\PY{p}{,} \PY{n+nv}{ptr}\PY{p}{)}

    \PY{n+nv}{ptr} \PY{o}{=}\PY{o}{\PYZgt{}} \PY{n+nv}{slv}\PY{p}{\PYZpc{}}\PY{n+nv}{courant}\PY{p}{(}\PY{l+m+mi}{0}\PY{p}{)} 
    \PY{n+nv}{ptr} \PY{o}{=} \PY{n+nv}{Cx}

    \PY{n+nv}{ptr} \PY{o}{=}\PY{o}{\PYZgt{}} \PY{n+nv}{slv}\PY{p}{\PYZpc{}}\PY{n+nv}{courant}\PY{p}{(}\PY{l+m+mi}{1}\PY{p}{)} 
    \PY{n+nv}{ptr} \PY{o}{=} \PY{n+nv}{Cy}

    \PY{k}{call }\PY{n+nv}{slv}\PY{p}{\PYZpc{}}\PY{n+nv}{solve}\PY{p}{(}\PY{n+nv}{nt}\PY{p}{)}
\PY{c}{!listing26}

    \PY{k}{block}
\PY{k}{      }\PY{k+kt}{logical} \PY{k+kd}{::} \PY{n+nv}{error} \PY{o}{=} \PY{n+nb}{.FALSE.}
      \PY{k+kt}{real}\PY{p}{(}\PY{n+nv}{real\PYZus{}t}\PY{p}{)}\PY{p}{,} \PY{k}{pointer} \PY{k+kd}{::} \PY{n+nv}{tmp}\PY{p}{(}\PY{p}{:}\PY{p}{,}\PY{p}{:}\PY{p}{)}
      \PY{k+kt}{real} \PY{k+kd}{::} \PY{n+nv}{diff}
      \PY{k+kt}{character} \PY{p}{(}\PY{n+nb}{len}\PY{o}{=}\PY{l+m+mi}{666}\PY{p}{)} \PY{k+kd}{::} \PY{n+nv}{fname}
      \PY{k}{allocate}\PY{p}{(}\PY{n+nv}{tmp}\PY{p}{(}\PY{n+nv}{nx}\PY{p}{,}\PY{n+nv}{ny}\PY{p}{)}\PY{p}{)}
      \PY{k}{call }\PY{n+nb}{get\PYZus{}command\PYZus{}argument}\PY{p}{(}\PY{l+m+mi}{8}\PY{p}{,} \PY{n+nv}{fname}\PY{p}{)}
      \PY{k}{call }\PY{n+nv}{read\PYZus{}file}\PY{p}{(}\PY{n+nv}{fname}\PY{p}{,} \PY{n+nv}{tmp}\PY{p}{)}
      \PY{n+nv}{diff} \PY{o}{=} \PY{n+nb}{maxval}\PY{p}{(}\PY{n+nb}{abs}\PY{p}{(}\PY{n+nv}{slv}\PY{p}{\PYZpc{}}\PY{n+nv}{state}\PY{p}{(}\PY{p}{)} \PY{o}{-} \PY{n+nv}{tmp}\PY{p}{)}\PY{p}{)}
      \PY{k}{print}\PY{o}{*}\PY{p}{,} \PY{l+s+s2}{"diff="}\PY{p}{,} \PY{n+nv}{diff}
      \PY{k}{if} \PY{p}{(}\PY{n+nv}{diff} \PY{o}{\PYZgt{}}\PY{o}{=} \PY{l+m+mf}{.5} \PY{o}{*} \PY{l+m+mi}{1}\PY{l+m+mf}{0.}\PY{o}{**}\PY{p}{(}\PY{o}{-}\PY{n+nv}{dec}\PY{p}{)}\PY{p}{)} \PY{n+nv}{error} \PY{o}{=} \PY{n+nb}{.TRUE.}
      \PY{k}{deallocate}\PY{p}{(}\PY{n+nv}{tmp}\PY{p}{)}
      \PY{k}{if} \PY{p}{(}\PY{n+nv}{error}\PY{p}{)} \PY{k}{then}
        \PY{c}{!print*, slv\PYZpc{}state()}
        \PY{k}{stop }\PY{l+m+mi}{1}
      \PY{k}{end }\PY{k}{if}
\PY{k}{    }\PY{k}{end }\PY{k}{block}
\PY{k}{  }\PY{k}{end }\PY{k}{block}

\PY{k}{  }\PY{k}{contains}

\PY{k}{  }\PY{k}{function }\PY{n+nv}{arg2int}\PY{p}{(}\PY{n+nv}{argno}\PY{p}{)} \PY{k}{result}\PY{p}{(}\PY{k}{return}\PY{p}{)}
    \PY{k}{implicit }\PY{k}{none}
\PY{k}{    }\PY{k+kt}{integer} \PY{k+kd}{::} \PY{n+nv}{argno}\PY{p}{,} \PY{k}{return}
\PY{k}{    }\PY{k+kt}{character} \PY{p}{(}\PY{n+nb}{len}\PY{o}{=}\PY{l+m+mi}{666}\PY{p}{)} \PY{k+kd}{::} \PY{n+nv}{tmp}
    \PY{k}{call }\PY{n+nb}{get\PYZus{}command\PYZus{}argument}\PY{p}{(}\PY{n+nv}{argno}\PY{p}{,} \PY{n+nv}{tmp}\PY{p}{)}
    \PY{k}{read}\PY{p}{(}\PY{n+nv}{tmp}\PY{p}{,}\PY{o}{*}\PY{p}{)}\PY{k}{return}
\PY{k}{  }\PY{k}{end }\PY{k}{function}

\PY{k}{  }\PY{k}{function }\PY{n+nv}{arg2real}\PY{p}{(}\PY{n+nv}{argno}\PY{p}{)} \PY{k}{result}\PY{p}{(}\PY{k}{return}\PY{p}{)}
    \PY{k}{implicit }\PY{k}{none}
\PY{k}{    }\PY{k+kt}{integer} \PY{k+kd}{::} \PY{n+nv}{argno}
    \PY{k+kt}{real}\PY{p}{(}\PY{n+nv}{real\PYZus{}t}\PY{p}{)} \PY{k+kd}{::} \PY{k}{return}
\PY{k}{    }\PY{k+kt}{character} \PY{p}{(}\PY{n+nb}{len}\PY{o}{=}\PY{l+m+mi}{666}\PY{p}{)} \PY{k+kd}{::} \PY{n+nv}{tmp}
    \PY{k}{call }\PY{n+nb}{get\PYZus{}command\PYZus{}argument}\PY{p}{(}\PY{n+nv}{argno}\PY{p}{,} \PY{n+nv}{tmp}\PY{p}{)}
    \PY{k}{read}\PY{p}{(}\PY{n+nv}{tmp}\PY{p}{,}\PY{o}{*}\PY{p}{)}\PY{k}{return}
\PY{k}{  }\PY{k}{end }\PY{k}{function}

\PY{k}{  }\PY{k}{subroutine }\PY{n+nv}{read\PYZus{}file}\PY{p}{(}\PY{n+nv}{fname}\PY{p}{,} \PY{k}{array}\PY{p}{)}
    \PY{k+kt}{character} \PY{p}{(}\PY{n+nb}{len}\PY{o}{=}\PY{o}{*}\PY{p}{)}\PY{p}{,} \PY{k}{intent} \PY{p}{(}\PY{n+nv}{in}\PY{p}{)} \PY{k+kd}{::} \PY{n+nv}{fname}
    \PY{k+kt}{real}\PY{p}{(}\PY{n+nv}{real\PYZus{}t}\PY{p}{)}\PY{p}{,} \PY{k}{pointer} \PY{k+kd}{::} \PY{k}{array}\PY{p}{(}\PY{p}{:}\PY{p}{,}\PY{p}{:}\PY{p}{)}
    \PY{k+kt}{integer} \PY{k+kd}{::} \PY{n+nv}{un}
    \PY{k}{open}\PY{p}{(}\PY{n+nv}{newunit}\PY{o}{=}\PY{n+nv}{un}\PY{p}{,} \PY{n+nv}{file}\PY{o}{=}\PY{n+nv}{fname}\PY{p}{,} \PY{n+nv}{status}\PY{o}{=}\PY{l+s+s1}{'old'}\PY{p}{,} \PY{n+nv}{action}\PY{o}{=}\PY{l+s+s1}{'read'} \PY{p}{)}
    \PY{k}{block    }
\PY{k}{      }\PY{k+kt}{integer} \PY{k+kd}{::} \PY{n+nv}{i}\PY{p}{,} \PY{n+nv}{j}
      \PY{k}{do }\PY{n+nv}{i}\PY{o}{=}\PY{l+m+mi}{1}\PY{p}{,} \PY{n+nv}{size}\PY{p}{(}\PY{k}{array}\PY{p}{,} \PY{l+m+mi}{1}\PY{p}{)}
        \PY{k}{read}\PY{p}{(}\PY{n+nv}{un}\PY{p}{,} \PY{o}{*}\PY{p}{)} \PY{p}{(}\PY{k}{array} \PY{p}{(}\PY{n+nv}{i}\PY{p}{,} \PY{n+nv}{j}\PY{p}{)}\PY{p}{,} \PY{n+nv}{j}\PY{o}{=}\PY{l+m+mi}{1}\PY{p}{,} \PY{n+nv}{size}\PY{p}{(}\PY{k}{array}\PY{p}{,} \PY{l+m+mi}{2}\PY{p}{)}\PY{p}{)}
      \PY{k}{end }\PY{k}{do}
\PY{k}{    }\PY{k}{end }\PY{k}{block}
\PY{k}{    }\PY{k}{close} \PY{p}{(}\PY{n+nv}{un}\PY{p}{)}
   \PY{k}{end }\PY{k}{subroutine }\PY{n+nv}{read\PYZus{}file}
\PY{k}{end }\PY{k}{program}
\end{Verbatim}
  \setcounter{linenofor}{\value{FancyVerbLine}}%

  \addtocounter{lstnofor}{1}%
  \renewcommand*\FancyVerbStartString{\PY{c}{!listing25}}
  \renewcommand*\FancyVerbStopString{\PY{c}{!listing26}}
  \setcounter{FancyVerbLine}{\thelinenofor}%
  \fvset{label={listing~F.\thelstnofor~(Fortran)},rulecolor=\color{red},stepnumber=1}%
  \setcounter{linenofor}{\value{FancyVerbLine}}%

  \bibliographystyle{model1-num-names}
  \bibliography{paper}
  
\end{document}